\documentclass[aps,prl,floats,floatfix,twocolumn,fleqn,linenumbers]{revtex4-1}

\RequirePackage{lineno} 
\renewcommand\thelinenumber{}

\usepackage{graphicx}
\usepackage{epstopdf}
\usepackage{overpic}
\graphicspath{{./Figs/}} 

\usepackage{latexsym}
\usepackage{amsmath}
\usepackage{amssymb}
\usepackage{bm}
\usepackage{wasysym}

\usepackage{ifthen}
\usepackage{color}
\usepackage[colorlinks,allcolors=blue,bookmarks=true]{hyperref}

\newcommand{\amatrix}[1]{\begin{matrix} #1 \end{matrix}} 
\newcommand{\eexp}[1]{\mathrm{e}^{#1}}

\newcommand{\braket}[1]{\left\langle #1 \right\rangle }

\newcommand{\Braket}[2]{\left\langle #1 \middle| #2 \right\rangle}

\newcommand{\beq}{\begin{eqnarray}}
\newcommand{\eeq}{\end{eqnarray}}

\newcommand{\hide}[1]{}  %{{\textcolor{red}{[hide]}}}
\newcommand{\rmrk}[1]{#1}

\newcommand{\Eq}[1]{\textcolor{blue}{{equation}\!~(\ref{#1})}} 
 
\newcommand{\Fig}[1] {{\textcolor{blue}{Fig.}}~\!\!\ref{#1}}
\newcommand{\sect}[1]{{\bf #1.-- }}

\newcommand{\Cn}[1]{\begin{center} #1 \end{center}}
\newcommand{\hrefl}[2]{\href{#2}{(#1)}}
\renewcommand{\thesection}{\arabic{section}}
\renewcommand{\thesubsection}{\arabic{subsection}}

\newcommand{\sectA}[1]
{
\addtocounter{section}{1}
\setcounter{subsection}{0}
\ \\
\pdfbookmark[1]{\thesection. \ #1}{sect.\thesection}
{\Large\bf $=\!=\!=\!=\!=\!=\;$ [\thesection] \ #1}
\nopagebreak
\vspace*{3mm}
}
\renewcommand{\section}{\sectA}

\newcommand{\sectB}[1]
{
\addtocounter{subsection}{1}
\ \\
\pdfbookmark[2]{\ \ \ \ \thesection.\thesubsection. \ #1}{subsect.\thesection.\thesubsection}
{\bf $=\!=\!=\!=\!=\!=\;$ [\thesection.\thesubsection] \ #1}
\nopagebreak
}
\renewcommand{\subsection}{\sectB}

\begin{document}

\title{Quasistatic transfer protocols for atomtronic superfluid circuits}

\author{Yehoshua Winsten, Doron Cohen}

\affiliation{
\mbox{Department of Physics, Ben-Gurion University of the Negev, Beer-Sheva 84105, Israel} 
}

\maketitle

%%%%%%%%%%%%%%%%%%%%%%%%%%%%%%%%%%%%%%%%%%%%%%%%%%%%%%%%%%%%%%%%%%%%%%%%%%%%%%%%%%%%%%%%%%
%%%%%%%%%%%%%%%%%%%%%%%%%%%%%%%%%%%%%%%%%%%%%%%%%%%%%%%%%%%%%%%%%%%%%%%%%%%%%%%%%%%%%%%%%%
%\vspace*{5mm}
\noindent {\Large\bf\textsl{Abstract}} 
\vspace*{1mm}

Quasi-static protocols for systems that feature a mixed phase-space with both chaos and quasi-regular regions are beyond the standard paradigm of adiabatic processes. We focus on a many-body system of atoms that are described by the Bose-Hubbard Hamiltonian, specifically a circuit that consists of bosonic sites. We consider a sweep process: slow variation of the rotation frequency of the device (time dependent Sagnac phase). The parametric variation of phase-space topology implies that the quasi-static limit is {\em irreversible}. Detailed analysis is essential in order to determine the outcome of such transfer protocol, and its efficiency.

%%%%%%%%%%%%%%%%%%%%%%%%%%%%%%%%%%%%%%%%%%%%%%%%%%%%%%%%%%%%%%%%%%%%%%%%%%%%%%%%%%%%%%%%%%
%%%%%%%%%%%%%%%%%%%%%%%%%%%%%%%%%%%%%%%%%%%%%%%%%%%%%%%%%%%%%%%%%%%%%%%%%%%%%%%%%%%%%%%%%%
\vspace*{5mm}
\noindent {\Large\bf\textsl{Introduction}} 
\vspace*{1mm}

Considering a closed Hamiltonian driven system, such as a particle in a box with moving wall (aka the piston paradigm), the common claim in Statistical Mechanics textbooks is that quasi-static (QS) processes are adiabatic, with vanishing dissipation in this limit, which implies thermodynamic reversibility. Indeed this claim can be \rmrk{established} for an {\em integrable} system by recognizing that the action-variables are adiabatic invariants \cite{Landau}. Also the other extreme, of a slowly driven completely {\em chaotic} system, has been addressed \cite{Ott1,Ott2,Ott3}, leading to the mesoscopic version of the Kubo linear-response result and the associated fluctuation-dissipation phenomenology \cite{Wilkinson1,Wilkinson2,crs,frc}. But generic systems are neither integrable nor completely chaotic. Rather they have {\em mixed phase space}. For such system the adiabatic picture fails miserly \cite{Kedar1,Kedar2,apc,lbt}, because the variation of the control parameter is associated with structural changes in phase space topology: tori merge into chaos, and new sets of tori are formed later on. This can be regarded as the higher-dimensional version of separatrix crossing   \cite{Kruskal,Neishtadt1,Timofeev,Henrard,Tennyson,Hannay,Cary,Neishtadt2,Elskens,Anglin,Neishtadt3}, where the so-called Kruskal-Neishtadt-Henrard theorem is followed.

In the present work we consider the implications of having mixed phase space with regard to quasi-static transfer protocols (QSTP). Specifically we focus on Bose-Hubbard circuits , and ask what is the outcome of a QS process whose aim is to transfer particles coherently from one orbital to another orbital. 
Systems that are described by \rmrk{the Bose-Hubbard Hamiltonian (BHH)} are of major interest both theoretically and experimentally \cite{Oberthaler,Steinhauer,exprBHH1,exprBHH2}. The simplest configuration is the BHH dimer (two sites), aka the Bosonic Josephson Junction (BJJ), see \cite{csd} and references therein. More generally there is an interest in lattice ring circuits that can serve as a SQUID or as a useful Qubit device \cite{Amico,Paraoanu,Hallwood,sfr}. The hope is that coherent operation might be feasible for BHH configuration with few sites, as already established for protocols that involve two sites (BJJ). \rmrk{The most promising configuration is naturally the 3-site trimer  
\cite{ref12,trimer2,trimer3,trimer4,trimer6,trimer15,trimer7,trimer19,trimerSREP1,trimer20,trimer18,trimer12,trimer13,trimerSREP2,gallemi,sfs,sfc,sfa,bhm}.}
For the analysis of such circuit one has to confront the handling of an underlying mixed phase space \cite{KolovskyReview,sfc,sfa}. 
In particular the implications of mixed phase-space on the stability of superflow has been explored in Ref.\cite{sfc,sfa,bhm}.

Striking forms of irreversibility can be observed in hysteresis experiments with ultracold atoms, both is double well geometry \cite{exprDimerHys} and in ring geometry \cite{exprRingRev,exprRingNIST}. \rmrk{For related theoretical studies see for example \cite{Swallow1,Swallow2,Swallow3,Swallow4,Swallow5}, where the emphasis is mainly on the parametric bifurcations of fixed points in phase space (notably the so-called swallow-tail loops). }
More recently the effect of {\em chaos} has been taken into account while studying the efficiency of a nonlinear stimulated Raman adiabatic passage \cite{apc}; and the Hamiltonian hysteresis that follows the reversal of the driving scheme \cite{lbt}.

\rmrk{Our interest in QSTP is motivated by hysteresis experiments with atomtronic superfluid circuits, as in \cite{exprRingNIST}.} 
Namely, we consider the following protocol for a ring-shaped circuit: 
(1) Initially, at the preparation stage, all the particles are condensed into the lowest momentum orbital that has a zero winding number; (2) The rotation frequency $\Phi$ of the ring is gradually changed, aka sweep process; (3) The final state of the system is probed, and the momentum distribution is measured. One possibility would be to find that all the particles are still condensed in a single orbital, possibly with a different winding number. This would be the case for a strictly quantum-adiabatic process, for which the system follows the ground state (GS), namely ${E(t) \sim E_{\text{GS}}(\Phi(t))}$. This would be also the case in the presence of a bath that induces relaxation towards the instantaneous GS. But such scenarios are not realistic because they require extremely slow sweep, and because we would not like to expose the system to external dissipation. We therefore ask what would be the result of such protocol for an isolated  system that undergoes a realistic slow sweep process. This is precisely the regime where a semiclassical perspective is most effective \cite{KolovskyReview,sfc,sfa}. The condensate, which is a many-body coherent state, is represented by a Gaussian-like distribution in phase space. At the preparation stage this {\em cloud} of points is located at the minimum of the potential. This minimum is a stationary point (SP) of the Hamiltonian. We ask what is the fate of the evolving cloud at the end of the sweep? Is it going to ergodize, or is it going to maintain some coherence?  In a larger context we are looking for a theory for the design of QSTP.

%%%%%%%%%%%%%%%%%%%%%%%%%%%%%%%%%%%
% orbitals
\begin{figure}[b]
\includegraphics[height=2cm]{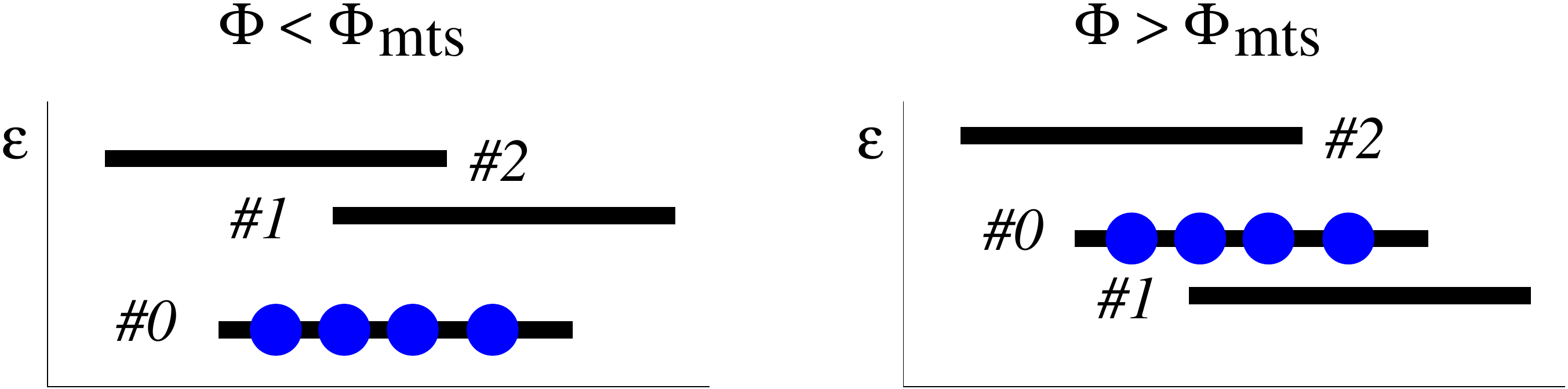}
\caption{{\bf \rmrk{Orbital occupation.}} 
For the purpose of illustration we consider a ring with $N{=}4$ particles.
The orbitals are represented by horizontal lines 
(the horizontal shifts hints the sign of the momentum).     
Initially (left panel) the particles are condensed in the \#0 momentum orbital. 
As $\Phi$ is increased beyond  $\Phi_{\text{mts}}$ this configuration becomes metastable (right). 
We ask what is the moment when the \#0 orbital is depleted, 
and what is the final distribution of the particles.
The $N{=}4$ system has 15 energy levels that corresponds to the different 
possibilities to distribute the particles between the orbitals. 
In the presence of non-zero interaction those levels are partially mixed.  
}	
\label{fRingOrbitals}
\end{figure}
%%%%%%%%%%%%%%%%%%%%%%%%%%%%%%%%%%%

% M vs Phi(t) 
% E vs Phi(t)
\begin{figure*}
%\centering	
\includegraphics[width=6cm]{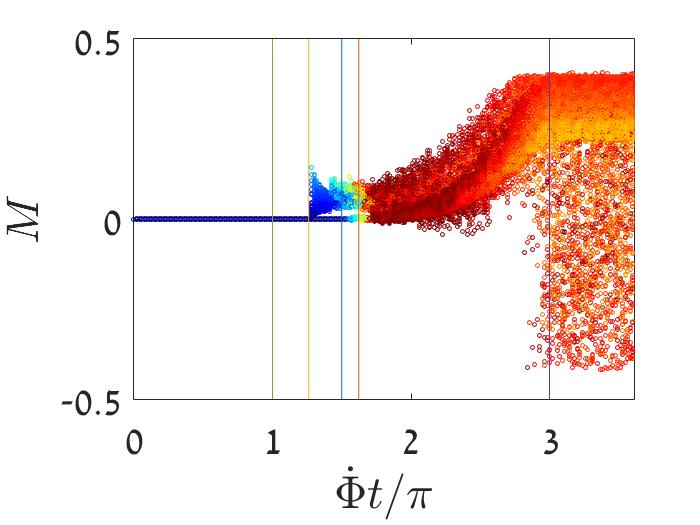}
\includegraphics[width=6cm]{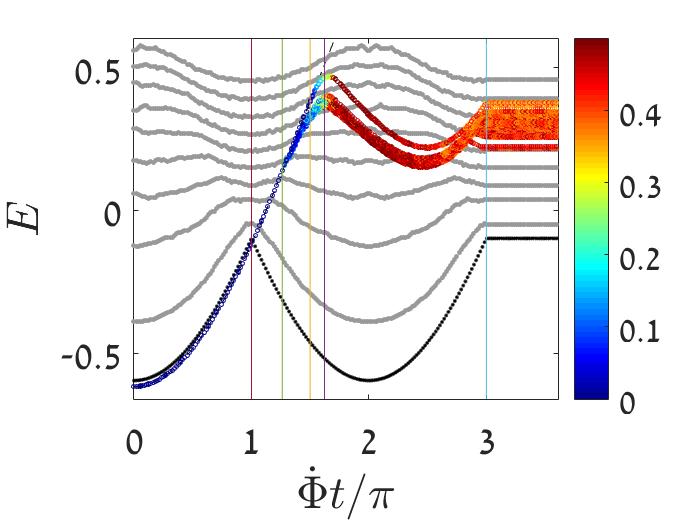} \\
%
%\includegraphics[width=6cm]{MasTime2-3--8000}
%\includegraphics[width=6cm]{EasTime2-3--8000}
%
%\setlength{\unitlength}{1cm}
%\begin{picture}(6,5)(0,0)
%\put(0,0){\includegraphics[width=6cm]{MasTime2-3--6000}}
%\put(1.55,1.1){\includegraphics[height=1.4cm,width=3.25cm]{IasTime2-3--6000}}  
%\end{picture} 
\includegraphics[width=6cm]{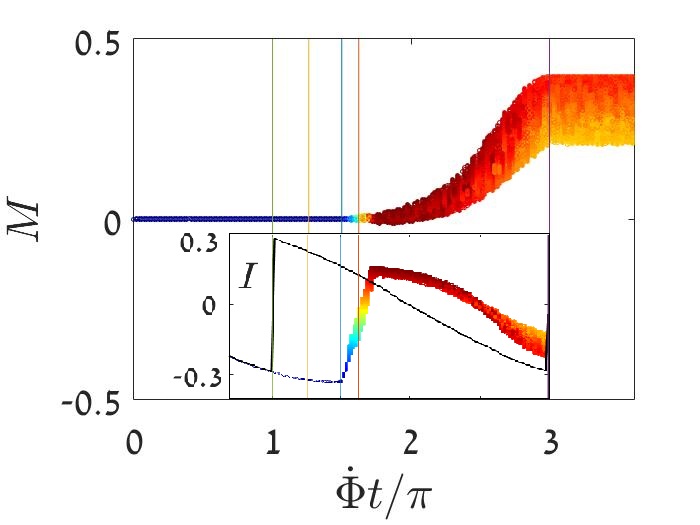}
\includegraphics[width=6cm]{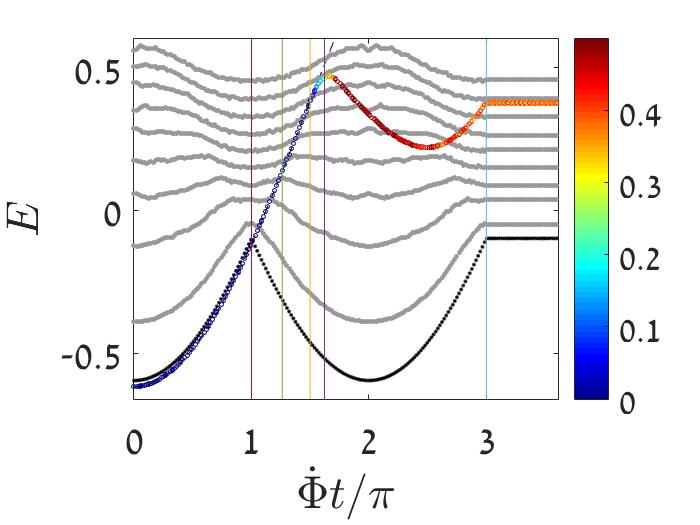} \\
\includegraphics[width=6cm]{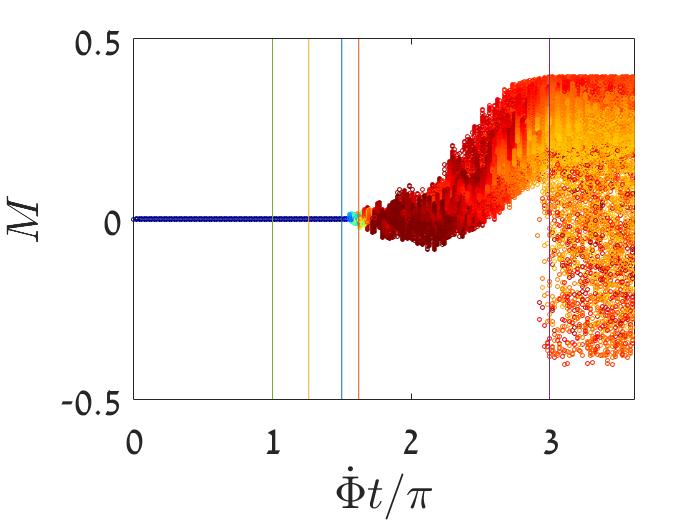}
\includegraphics[width=6cm]{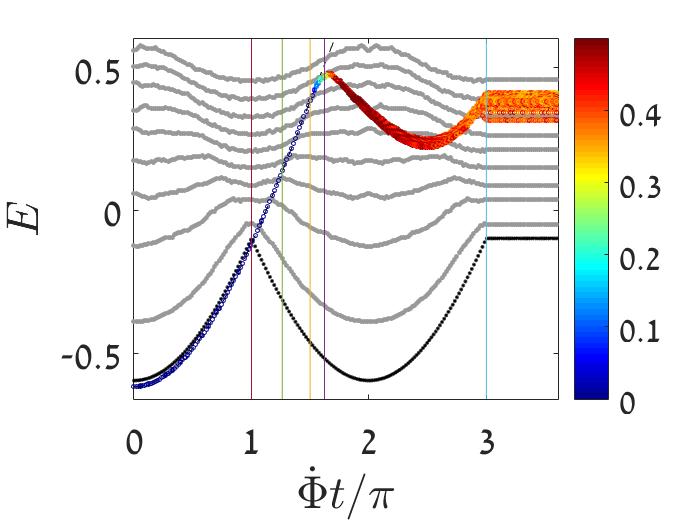}
\caption{ \label{fMaEvsTime}
{\bf Semiclassical simulation of a sweep process.}  
Here and below we consider a 3-site ring. 
The initial condensate is represented by a cloud of radius $R {=} 0.0001$ at $n{=}0$.  
\textbf{Left}: The $(n,M)$ coordinates of the evolving trajectories are presented as a function of time. Both coordinates are normalized (${n:=n/N, M:=M/N}$). 
The $n$ values are color coded such that blue corresponds to $n{=}0$ and red to total depletion.   
\textbf{Right}: The energy $E$ of the  evolving points as function of time.
The dotted line is the ground state energy $E_{\text{GS}}$, 
and the dashed line is the condensate energy $E_0$. 
The other lines in the background are subset of adiabatic $E_n$ curves (see text).  
\textbf{Inset} (second row): The current~$I$ that flows in the ring as a function of time.
\textbf{Parameters}: The interaction is $u{=}2.3$, 
and the associated vertical lines are from left to right $\Phi_{\text{mts}}{=}\pi$ 
and $\Phi_{\text{stb}}{=}1.26\pi$ and $\Phi_{\text{dyn}}{=}(3/2)\pi$ and $\Phi_{\text{swp}}{=}1.62\pi$. 
The units of time have been chosen such that $K{=}1$. 
Each row is for a different sweep rate. 
From up to down we have $\dot{\Phi}=3\pi \cdot 10^{-4}$ (slow) 
and $\dot{\Phi}=5\pi \cdot 10^{-4}$ (optimal) 
and $\dot{\Phi}=3\pi \cdot 10^{-3}$ (faster). 
}
\end{figure*}

%%%%%%%%%%%%%%%%%%%%%%%%%%%%%%%%%%%%%%%%%%%%%%%%%%%%%%%%%%%%%%%%%%%%%%%%%%%%%%%%%%%%%%%%%%%%%
\sect{Outline}
We present the model Hamiltonian in terms of physically motivated coordinates, and display results of sweep simulations. Then we illuminate our findings by performing step-by step analysis of the energy landscape, 
and of the phase-space dynamics.

%%%%%%%%%%%%%%%%%%%%%%%%%%%%%%%%%%%%%%%%%%%%%%%%%%%%%%%%%%%%%%%%%%%%%%%%%%%%%%%%%%%%%%%%%%
%%%%%%%%%%%%%%%%%%%%%%%%%%%%%%%%%%%%%%%%%%%%%%%%%%%%%%%%%%%%%%%%%%%%%%%%%%%%%%%%%%%%%%%%%%
\vspace*{5mm}
\noindent {\Large\bf\textsl{Results}} 
\vspace*{1mm}

%%%%%%%%%%%%%%%%%%%%%%%%%%%%%%%%%%%%%%%%%%%%%%%%%%%%%%%%%%%%%%%%%%%%%%%%%%%%%%%%%%%%%%%%%%%%%
\sect{The model}
We consider a system with $N$ bosons in a 3-site ring.
The system is described by the Bose-Hubbard Hamiltonian \rmrk{[Methods]} 
with hopping frequency $K$ and on-site interaction $U$.
The sweep control-parameter is the Sagnac phase $\Phi$, 
which is proportional to the rotation frequency of the device: 
it can be regarded as the Aharonov-Bohm flux that is associated
with Coriolis field in the rotating frame~\cite{exprRingRev,exprRingNIST}.
There are 3 momentum orbitals $k=0,\pm 2\pi/3$.
Initially all the particles are condensed in $k{=}0$.
\rmrk{A caricature for the preparation is provided in \Fig{fRingOrbitals} (left panel).}

Following \cite{bhm} we define a depletion coordinate~$n$ 
and an imbalance coordinate~$M$, such that the 
occupations of the orbitals are $n_{0}=N{-}2n$, and $n_{\pm}=n{\pm}M$. 
The model Hamiltonian can be written in terms of ${(n,M)}$, 
and the conjugate phases ${(\varphi,\phi)}$. Namely \rmrk{[Methods]}:
\beq \label{eHfull}
\mathcal{H}(\varphi,n;\phi,M) = \mathcal{H}^{(0)}(\varphi,n;M) +  \left[\mathcal{H}^{(+)} + \mathcal{H}^{(-)} \right] \ \ \ 
\eeq
The first term $\mathcal{H}^{(0)}$ is an integrable piece of the Hamiltonian 
that has~$M$ as a constant of motion: 
\beq \nonumber
&& \mathcal{H}^{(0)}(\varphi,n;M) \ = \ 
E_{0} + \mathcal{E}_{\parallel} n  + \mathcal{E}_{\perp} M - \frac{U}{3}M^2   
\\ \label{eH0}
&& \ \ + \frac{2U}{3} (N-2n)\left[ \frac{3}{4}n + \sqrt{n^2-M^2} \cos(\varphi) \right] \ \ \  
\eeq  
while the additional terms induce resonances that spoil the integrability, and give rise to chaos: 
\beq \label{eHchaos}
\mathcal{H}^{(\pm)} = \frac{2U}{3} \sqrt{(N{-}2n) (n {\pm} M)} (n {\mp} M) \cos \left(\frac{3\phi {\mp} \varphi}{2} \right) \ \ \ \ 
\eeq  
The hopping frequency $K$ and the Sagnac phase $\Phi$ 
hide in the expression for the energy of the condensate, 
and in the detuning parameters:
\beq
\label{eE0} E_0 \ &=& \ -NK \cos{\left(\frac{\Phi}{3}\right)} +  \frac{1}{6}UN^2 \\
\label{eEDn} \mathcal{E}_{\parallel} \ &=& \ 3K\cos{\left(\frac{\Phi}{3}\right)} +  \frac{1}{6}UN \\
\label{eEDM} \mathcal{E}_{\perp} \ &=& \ -\sqrt{3}K\sin{\left(\frac{\Phi}{3}\right)}
\eeq
\rmrk{We also note that the energies of the totally depleted states (${n=(N/2)}$) are  
\beq \label{eEMinfty} 
E_{\infty}(M) \ = \ E_0 + \mathcal{E}_{\parallel} \frac{N}{2}  + \mathcal{E}_{\perp} M - \frac{U}{3}M^2
\eeq
Note that the latter expression has zero contribution from the $\mathcal{H}^{(\pm)}$ terms. 
The {\em chaos} affects the pathway between the initial condensate at ${n{=}M{=}0}$, and the peripheral 
depleted states at ${n=(N/2)}$, but has only little effect on the gross features of the energy landscape.}

%%%%%%%%%%%%%%%%%%%%%%%%%%%%%%%%%%%%%%%%%%%%%%%%%%%%%%%%%%%%%%%%%%%%%%%%%%%%%%%%%%%%%%%%%%%%%
\sect{Metastability}
\rmrk{The central point in phase space ${n{=}M{=}0}$ is a stationary point (SP)
of the Hamiltonian for any $\Phi$, meaning that we have there ${\dot{n}=0}$.
But this does not mean that this SP is stable.          
As implied by the caricature of \Fig{fRingOrbitals}, 
the condensate at ${n{=}0}$ is no longer situate at the minimum 
of the energy landscape once ${ E_0 > \min\{ E_{\infty}(M) \} = E_{\infty}(N/2) }$.  
This leads to the threshold 
\beq \label{PhiMTS}
\Phi_{\text{mts}} \ \ = \ \ \pi
\eeq 
Once we cross ${\Phi_{\text{mts}}}$ the SP becomes a metastable minimum.  
Illustrations of the energy landscape for representative values of $\Phi$ can be found in \rmrk{SM}. 
In the subsequent paragraphs we shall discuss additional thresholds:
Once we cross ${\Phi_{\text{stb}}}$ the central SP becomes a saddle in the energy landscape. 
Once we cross ${\Phi_{\text{dyn}}}$ this saddle becomes dynamically unstable.
When ${ E_0 = E_{\infty}(0) }$ a dynamical corridor is opened between the central SP and the peripheral depleted states, 
leading to the identification of what we call swap transition at ${ \Phi = \Phi_{\text{swp}} }$.}

%%% define $u$
%%%%%%%%%%%%%%%%%%%%%%%%%%%%%%%%%%%%%%%%%%%%%%%%%%%%%%%%%%%%%%%%%%%%%%%%%%%%%%%%%%%%%%%%%%%%%
\sect{Semiclassics}
The classical (as opposed to semiclassical) treatment of the Hamiltonian is commonly 
termed Mean Field Theory (MFT). The evolving state is represented by a single point 
in phase space. We can scale the time such that ${t:= Kt}$,  
and the occupations such that ${n:= n/N}$. Then one finds that the dynamics 
is controlled by the dimensionless interaction parameter 
\beq
u \ \ = \ \ \frac{NU}{K}
\eeq
\rmrk{Upon quantization (aka ``second quantization") the scaled value 
of the Planck constant is ${\hbar=1/N}$, see e.g. \cite{KolovskyReview,sfc,sfa}. 
Quantum states can be represented in phase space by their Wigner function.  
In particular the initial coherent state at $n{=}0$ is represented 
in phase-space by a Gaussian-like distribution of radius ${R \sim 1/N }$.}

\rmrk{What we call ``semiclassical treatment" is far better and reliable
compared to MFT, and is commonly called Truncated Wigner Approximation (TWA). 
Within the framework of TWA the Moyal brackets are approximated by Poisson brackets, 
which means that the Wigner function is propagated by the classical equations of motion.} 

\rmrk{The TWA is very accurate as long as quantum tunneling is neglected. 
The tunneling amplitude scales as $\exp[-\text{Action}/\hbar]$, 
where ${\hbar=1/N}$. Therefore it is much slower compared with any classical process. 
Discussion of tunneling in the BHH context can be found in \cite{dimerSplit}, 
and later we demonstrate numerically that it can be neglected 
for a simulation with ${N=30}$ particles.}

%%%%%%%%%%%%%%%%%%%%%%%%%%%%%%%%%%%%%%%%%%%%%%%%%%%%
% Efficiency: <n> and <M> vs $dot{Phi}$ for misc $u$ 
\begin{figure}
\centering
\includegraphics[width=8cm]{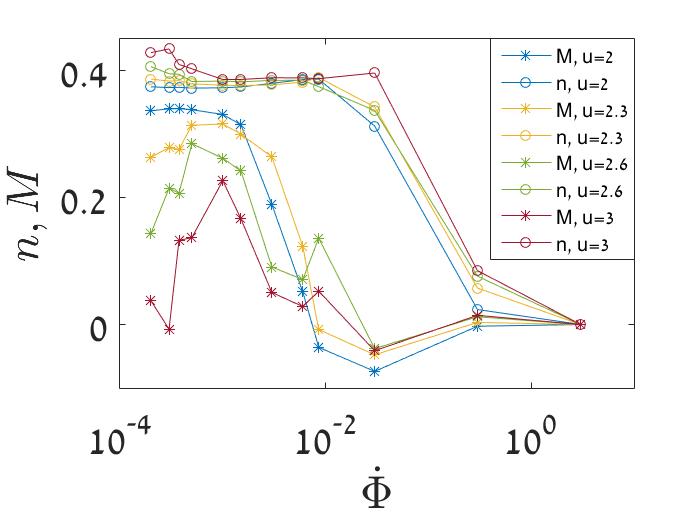}
\caption{ 
{\bf Efficiency of the sweep process.} 
The expectation values $\braket{n}$ and $\braket{M}$ at the end of the sweep process 
are plotted against $\dot{\Phi}$ for misc values of $u$. 
Note again that the coordinates are normalized (${n:=n/N, M:=M/N}$). 
The optimal sweep rate is determined by inspection of the maximum of~$\braket{M}$, 
which becomes prominent for large~$u$.  
}
\label{fEfficiency}
\end{figure}

%%%%%%%%%%%%%%%%%%%%%%%%%%%%%%%%%%%%%%%%%%%%%%%%%%%%%%%%%%%%%%%%%%%%%%%%%%%%%%%%%%%%%%%%%%%%%
%%%%%%%%%%%%%%%%%%%%%%%%%%%%%%%%%%%%%%%%%%%%%%%%%%%%%%%%%%%%%%%%%%%%%%%%%%%%%%%%%%%%%%%%%%%%%
\sect{Simulations}
We describe the results of semiclassical simulations. 
Detailed analysis will follow after that.      
The condensate preparation at $\Phi{=}0$ is represented by a Gaussian cloud of points in phase space, 
at the central SP ($n{=}0$).
The evolution of the cloud in a dynamical sweep simulation is demonstrated in \Fig{fMaEvsTime}. 
The color-code shows the evolution of the depletion coordinate ($n$), 
and the vertical position of the cloud points indicate  
the population imbalance $M$ (left panels), 
or the energy $E{=}\mathcal{H}$ (right panels), 
or the current $I = -{\partial \mathcal{H}} / {\partial \Phi}$ 
as a function of time (inset). For the latter we use 
the following expression in terms of~${(n,M)}$,
\beq \label{eq:I}
I \ = \ \left(n{-}\frac{N}{3}\right) K \sin{\frac{\Phi}{3}} \, +  \, \frac{M}{\sqrt{3}} K \cos{\frac{\Phi}{3}}
\eeq
Note that the cloud is a semiclassical representation of the evolving state. Accordingly, to get the expectation value of the energy or of the current, an average has to be taken over the ensemble of evolving trajectories. In \Fig{fMaEvsTime} the average is not taken in order to provide an insight for the dispersion as well.

The cloud follows the ground state energy $E_{\text{GS}}$ only up to $\Phi_{\text{mts}}$. 
Then it continues to follow the condensate energy $E_0$ during an additional time interval. 
The cloud {\em starts} spreading not before $\Phi_{\text{stb}}$, 
and not later than $\Phi_{\text{dyn}}$. 
The spreading is indicated by the departure of energy from $E_0$.  
The depletion of the condensate is indicated by 
the color that changes abruptly from blue ($n{=}0$) to red ($n{\sim}N/2$). 
It takes place during a distinct short time interval when $\Phi(t) \sim \Phi_{\text{swp}}$.   
The depletion stage is also clearly reflected as a jump in the current-versus-time plot.  
Finally, the subsequent evolution after the depletion does not follow any of the adiabatic $E_n$ curves, as discussed further below. 
 
We display in \Fig{fMaEvsTime} three representative simulations: 
very slow sweep (top row), {\em optimal} sweep rate (middle row), 
and faster sweep (lower row).  
The results for many such simulations are gathered in \Fig{fEfficiency}, 
where the dependence of $\braket{n}$ and $\braket{M}$ on the sweep rate 
is demonstrated for different values of the interaction~$u$.  
What we call optimal sweep rate provides the most coherent outcome (minimum dispersion). 
Contrary to the traditional dogma, it is not true that ``slower is better".

\begin{figure}
(a) \hspace*{5cm} \\ \includegraphics[width=7cm]{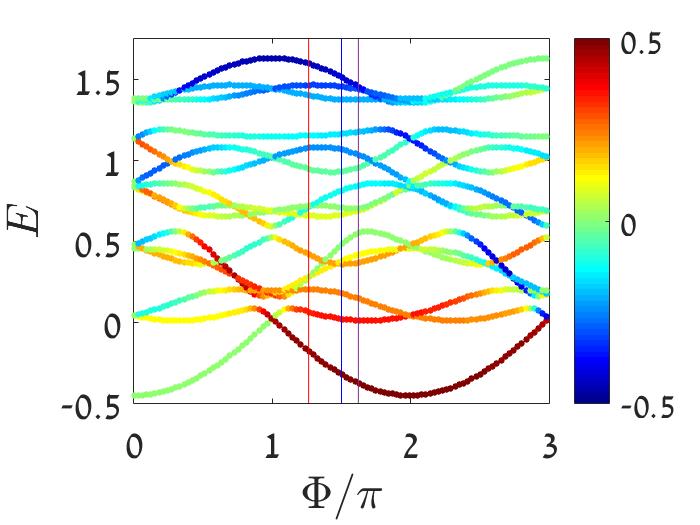}\\
(b) \hspace*{5cm} \\ \includegraphics[width=7cm]{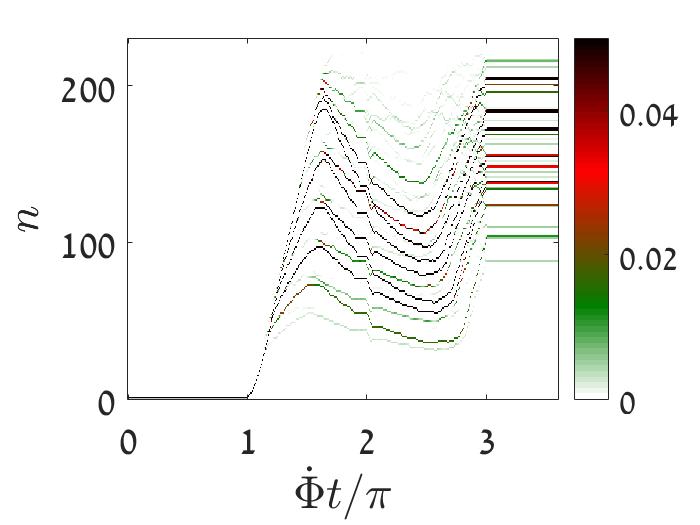}
\caption{{\bf Quantum spreading in few particle system.} 
Even for small number of particles the semiclassical perspective is useful.  
{\bf Upper panel:} The many body energy levels $E_n$ for $N{=}4$ particles in a 3-site ring as a function of $\Phi$ for $u{=}2.3$. The points are color-coded by the expectation value of $M$. 
{\bf Lower panel:} The quantum evolution of $N{=}30$ particle ring is imaged. 
Each row is the color-coded probability ${p_n=|\Braket{E_n}{\psi(t)}|^2}$ as a function of time.
For larger $N$ we expect a very good quantitative correspondence with the semiclassical simulations of \Fig{fMaEvsTime}.        
}	
\label{fEn}
\end{figure}

%%%%%%%%%%%%%%%%%%%%%%%%%%%%%%%%%%%%%%%%%%%%%%%%%%%%%%%%%%%%%%%%%%%%%%%%%%%%%%%%%%%%%%%%%%%%%
\sect{Adiabatic evolution}
It is illuminating to discuss the $\Phi$ dependence of the energy landscape using a quantum ``energy level" language. The parametric evolution of the many body eigen-energies is presented in \Fig{fEn}a. If the system were completely chaotic, then we could associate each $E_n$ with a micro-canonical energy surface that encloses a phase space volume 
\beq \label{eNcells}
n \ \ = \ \ \mathcal{N}(E) \ \ \ \text{[Planck cells]}.
\eeq
\rmrk{Here, for a given number of particles, we have a system with $d{=}2$ degrees of freedom, 
and $\mathcal{N}(E)$ is the $2d$ hyper-volume of ${\mathcal{H}<E}$ divided by $(2\pi\hbar)^d$.}  
Irrespective of chaos, a practical numerical procedure to find the phase-space volume is to invert the dependence $E=E_n$ where ${n=1,2,3...}$. The validity of this statement is implied by the Wigner-Wyle formalism. The representative $E_n$ curves in the background of \Fig{fMaEvsTime} have been calculated using this procedure with $N{=}30$.      

For an adiabatic sweep, the phase-space volume \Eq{eNcells} is the so-called adiabatic invariant \cite{Ott1,Ott2,Ott3}. This statement assumes a globally chaotic energy surface. In the classical context we say that during an adiabatic sweep the system stays in the same adiabatic energy surface. 
In the quantum context we say that the system stays in the same adiabatic energy level.

In a strictly quantum-adiabatic scenario, the system stays in its ground state with energy $E_{\text{GS}}(\Phi)$, and therefore the population is fully depleted from $k{=}0$ to the other orbitals. \rmrk{Such quantum adiabaticity cannot be observed for a realistic sweep rate, because it requires many-body tunneling from a metastable minimum of the energy landscape \cite{dimerSplit}.} Consequently, for large $N$, the semiclassical picture provides a sound approximation. \rmrk{In \Fig{fEn}b} we demonstrate that even a circuit with small number of particles ($N{=}30$) follows a semiclassical-like scenario. 

The semiclassical adiabatic scenario excludes the possibility of tunneling, and therefore can start only when ${\Phi(t) > \Phi_{\text{stb}}}$, namely, once the central SP becomes a saddle in the energy landscape. In order to determined $\Phi_{\text{stb}}$ we use the Bogolyubov procedure, which brings the Hamiltonian in the vicinity of the SP to a diagonalized form: 
\beq \label{BG}
\mathcal{H} \ \ \approx \ \ E_0 + \sum_q \omega_q c_q^{\dag} c_q
\eeq
Explicit results for the Bogolyubov frequencies are provided in the \rmrk{Methods} section.   
The SP becomes a saddle once the $\omega_q$ do not have the same sign.
This happens for $\Phi$ larger than 
\beq \label{PhiSTB}
\Phi_{\text{stb}} = \ \ 3 \arccos{\left(\frac{1}{6} \left(\sqrt{u^2+9}-u\right)\right)}
\eeq  
\rmrk{The topography at the vicinity of the central SP, 
once it becomes a saddle is as follows:     
it is still a minimum in the $M{=}0$ subspace, 
while away from $M{=}0$ the energy floor is lower 
(see \rmrk{SM} for plots of the $(M,E)$ energy landscape). 
Nevertheless, we see from the simulation of \Fig{fMaEvsTime} 
that spreading away from the central SP 
starts only at a later stage,   
whereafter the cloud departs the $E_0$ curve, 
neither follows any of the $E_n$ curves.}

%%%%%%%%%%%%%%%%%%%%%%%%%%%%%%%%%%%
% Energy levels for u=2.3
% Quantum simulation

\begin{figure}
(a) \hspace*{7cm}  \\ \vspace*{-4mm}
\includegraphics[width=6.6cm]{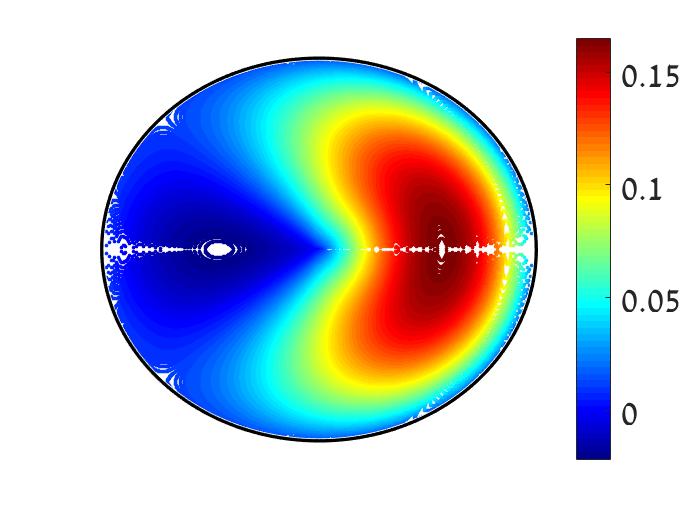} \\
(b) \hspace*{7cm}  \\ \vspace*{-4mm}
\includegraphics[width=6.6cm]{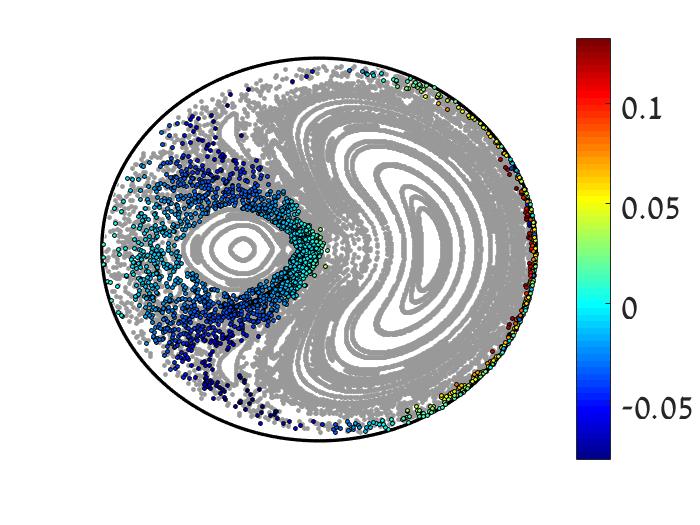} \\
(c) \hspace*{7cm}  \\ \vspace*{-4mm} \hspace*{10mm}
\begin{overpic}[width=7.5cm]{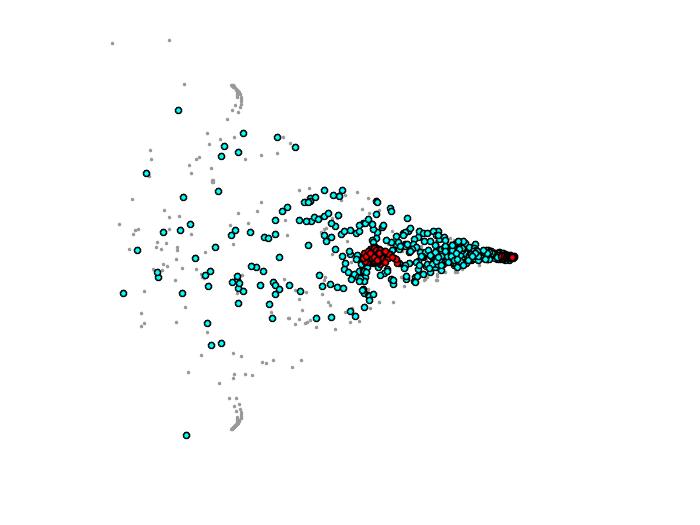} 
\put (60,45){\frame{{\includegraphics[scale=0.17]{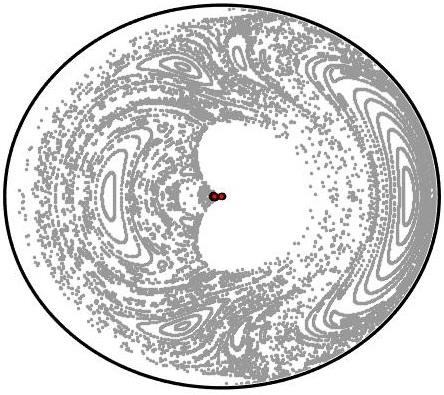}}}}
\put (60,13){\frame{{\includegraphics[scale=0.17]{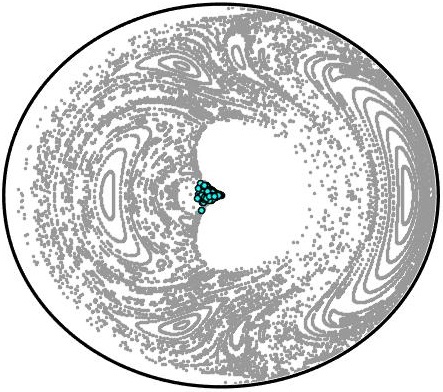}}}}
\end{overpic}
\caption{ \label{fPS}
{\bf Spreading and depletion.} 
{\bf (a)}~Image of $M$ for the phase space points 
of the energy surface $\mathcal{H}^{(0)}(\varphi,n;M)=E_0$.
The interaction is $u{=}2.3$ and ${\Phi{=}1.61\pi \sim \Phi_{\text{swp}}}$.   
{\bf (b)}~Poincare section for the same $\Phi$ at the same energy (gray trajectories),
and a spreading cloud (colored trajectories) following a {\em quench} to this $\Phi$ value. 
\rmrk{The initial cloud is the preparation at ${n{=}0}$. 
It spreads away from the central SP, and stretches along the 
chaotic corridor.} Its points are color-coded by~$M$.
{\bf (c)}~The spreading cloud in the {\em sweep} simulation.
Upper inset (red points): 
The sweep rate is ${\dot{\Phi}{=}3\pi \cdot 10^{-4}}$ (slow). 
The snapshot is taken at ${\Phi \sim \Phi_{\text{stb}}}$. 
An inner piece of the cloud is still locked in the tiny $n{=}0$ stability island, 
and therefore has energy close to~$E_0$.    
An outer piece of the cloud was formed due to very slow spreading in the chaotic corridor, 
and therefore has lower energy.  
The Poincare section at the background is adjusted to this lower energy.
Lower inset (blue points): The further evolution of the same cloud after we stop the 
sweep at ${\Phi{=}\Phi_{\text{stb}}}$ and wait to see further ergodization in the chaotic strip. 
The upper inset would look like that if the sweep were much slower.
Main panel: Zoom that displays the red and the blue clouds of the insets.    
}  
\end{figure}

%%%%%%%%%%%%%%%%%%%%%%%%%%%%%%%%%%%%%%%%%%%%%%%%%%%%%%%%%%%%%%%%%%%%%%%%%%%%%%%%%%%%%%%%%%%%%
\sect{Quench-related spreading}
Let us consider first the simpler scenario of preparing a cloud at ${n=0}$, which is the $\Phi{=}0$ ground state, and then evolving it with $\mathcal{H}(\Phi \ne 0)$, aka a quench process.  \rmrk{The SP for ${t>0}$ (after the quench) is dynamically unstable if the  Bogolyubov frequencies become complex.}  
This happens (see \rmrk{Methods}) for $\Phi$ larger than 
\beq \label{PhiDYN}
\Phi_{\text{dyn}} \ \ = \ \ \frac{3}{2}\pi
\eeq 
After the quench the cloud spreads away from $n{=}0$   
\rmrk{in the landscape that is described by \Fig{fPS}a,} 
as illustrated in \Fig{fPS}b. 
The Poincare section there shows that the stability island is taken-over by a chaotic strip. 
The points of the spreading cloud are colored. The other trajectories, 
that do not belong to the cloud, are not color-coded. 
If they were color-coded, one would see that for quasi-regular trajectories $M$ is approximately a constant of motion.

%%%%%%%%%%%%%%%%%%%%%%%%%%%%%%%%%%%
% M contors for E=E0
% Poincare sections for E=E0

%%%%%%%%%%%%%%%%%%%%%%%%%%%%%%%%%%%
% M contors for E=E0
% Poincare sections for E=E0
\begin{figure}

\noindent
\includegraphics[width=3cm]{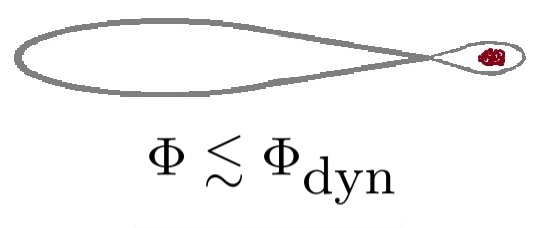}
\ \ \ \ \ 
\includegraphics[width=3cm]{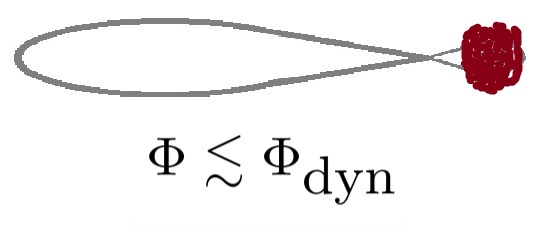}
\\ \ \ \ 
\includegraphics[width=3.5cm]{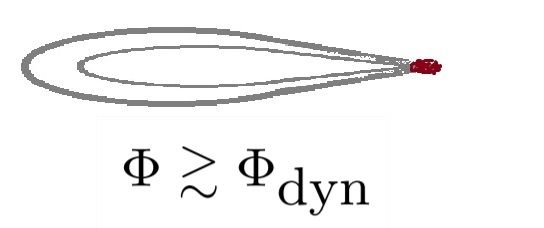}
\ \ \ 
\includegraphics[width=3.5cm]{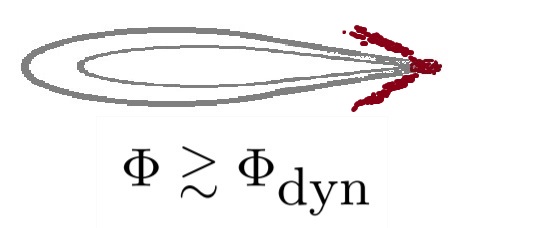}
\\
\includegraphics[width=4cm]{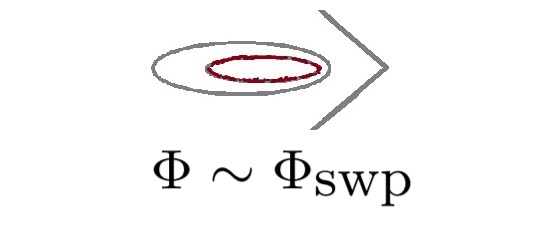}
\ \ \ 
\includegraphics[width=4cm]{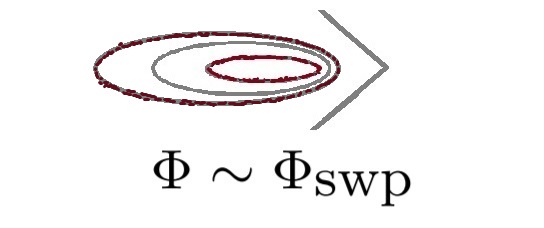}
\caption{ \label{fDYN}
\rmrk{{\bf The spreading mechanism.}} 
The dynamics of \Fig{fPS} is caricatured for an optimal sweep (left panels) 
and for a slow sweep (right panels). The panels are ordered by time (from top to bottom). 
For an optimal sweep, chaos has no time to induce spreading, 
therefore, even if the cloud is larger (not displayed) the spreading process looks the same. 
For a slow sweep the outer part of the cloud has the time to spread way 
from the center along the chaotic strip. This chaotic spreading is initiated 
in the range ${ [\Phi_{\text{stb}}, \Phi_{\text{dyn}}] }$, 
while the former takes place after ${\Phi_{\text{dyn}} }$,  
as clearly observed in the upper left panel of \Fig{fMaEvsTime}. 
}       
\end{figure}

%%%%%%%%%%%%%%%%%%%%%%%%%%%%%%%%%%%
% Dynamics: snapshots of the evolving cloud in (E,M,n) space 
\begin{figure*}
\centering
\begin{minipage}{8.5cm}
	\begin{overpic}[width=4cm]{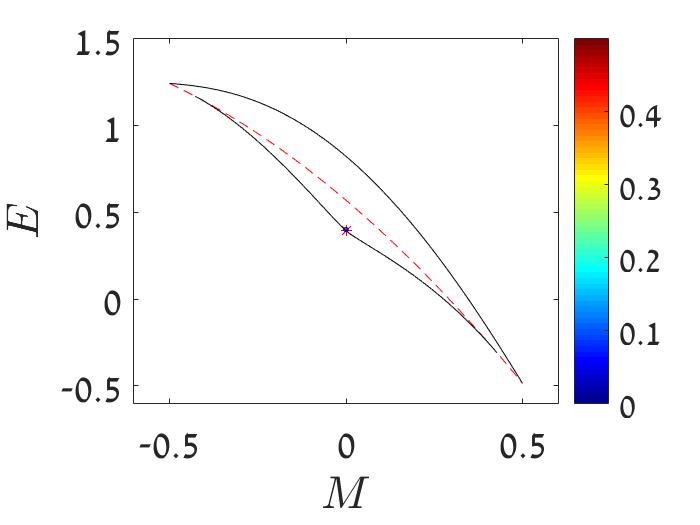} \put (55,50){\frame{{\includegraphics[scale=.095]{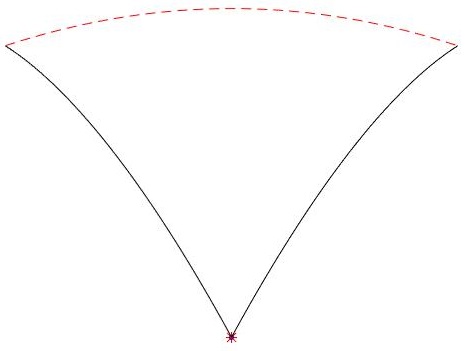}}}}\end{overpic}
\includegraphics[width=4cm]{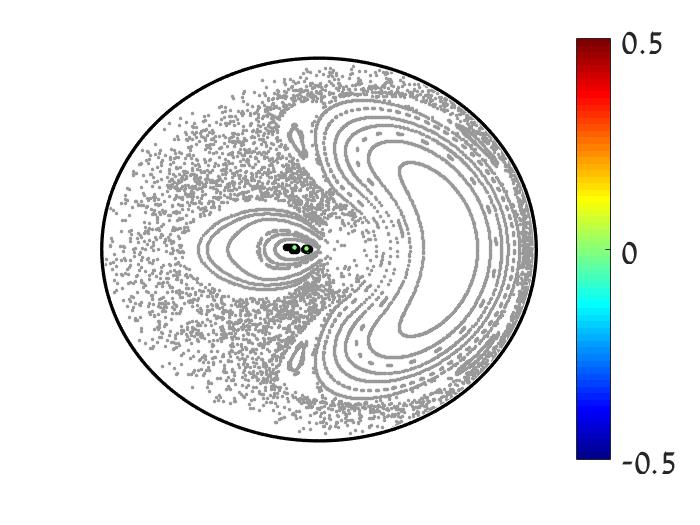}
\\
\includegraphics[width=4cm]{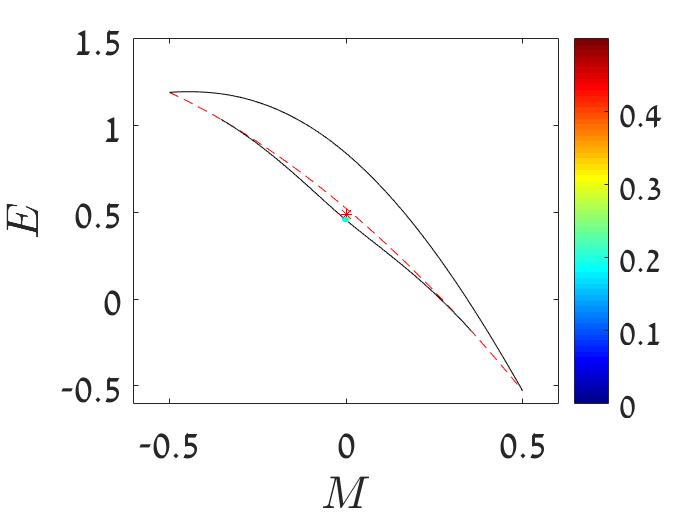}
\includegraphics[width=4cm]{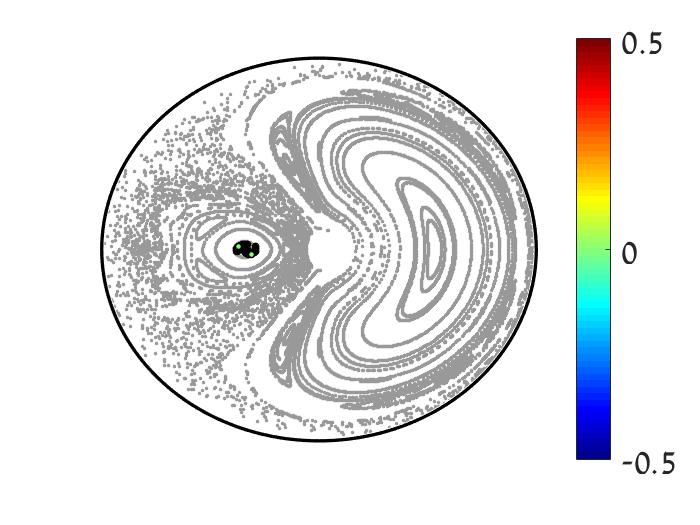}
\\
\includegraphics[width=4cm]{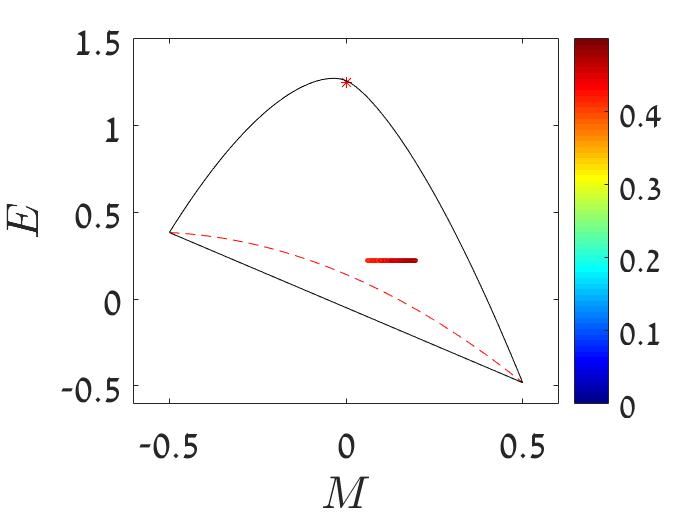}
\includegraphics[width=4cm]{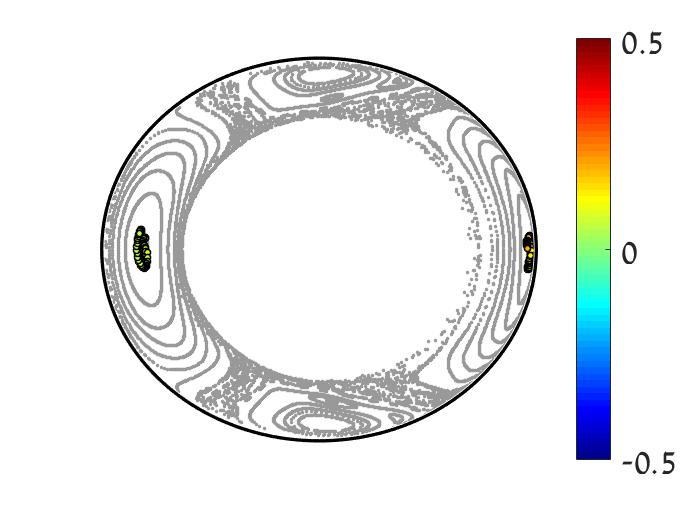}
\\
\includegraphics[width=4cm]{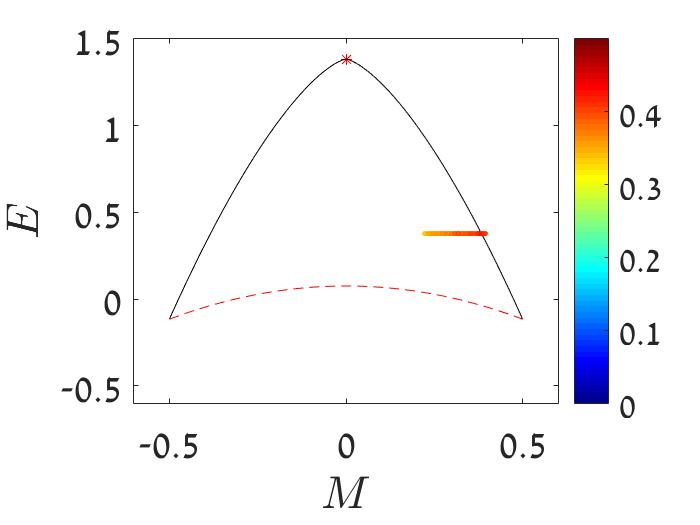}
\includegraphics[width=4cm]{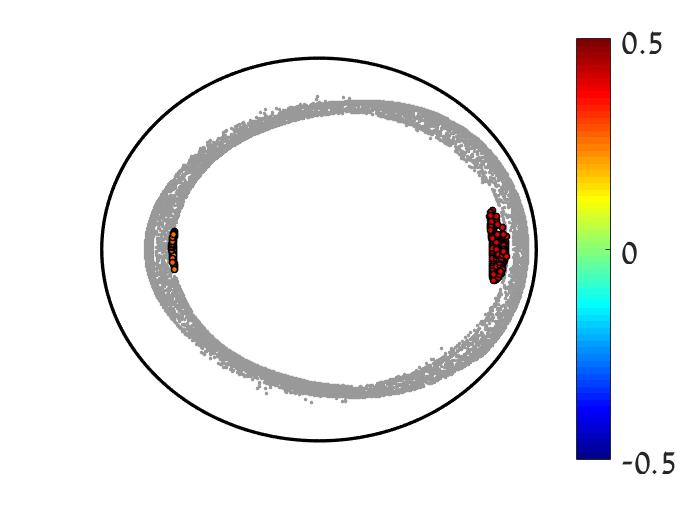}
\\
\includegraphics[width=7cm]{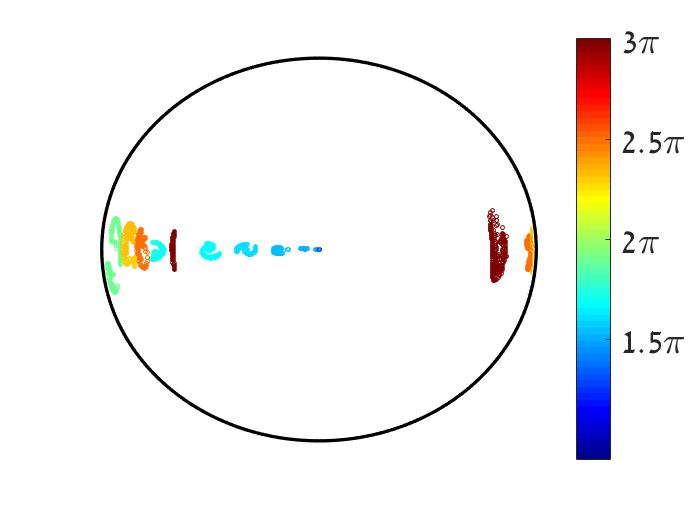}
\end{minipage}
\begin{minipage}{8.5cm}
	\begin{overpic}[width=4cm]{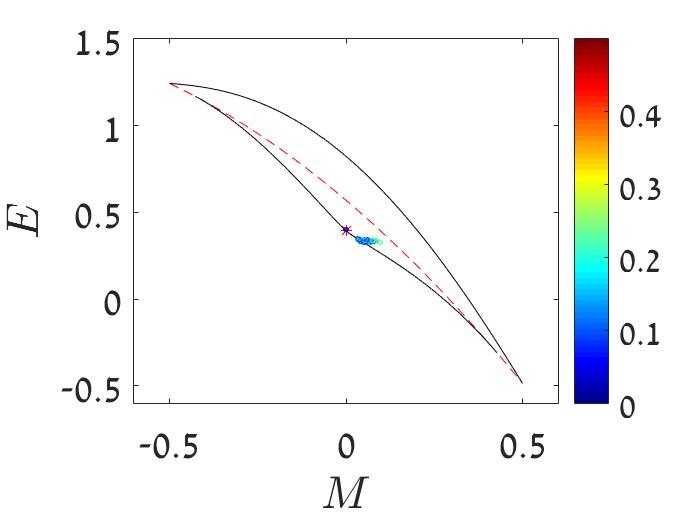} \put (55,50){\frame{{\includegraphics[scale=.095]{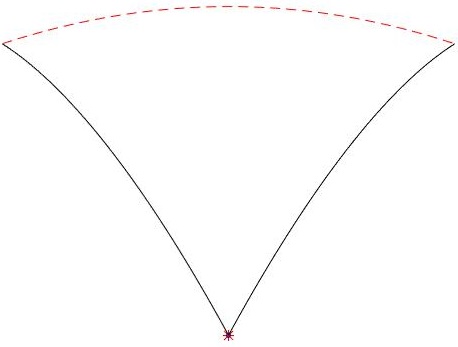}}}}\end{overpic}
\includegraphics[width=4cm]{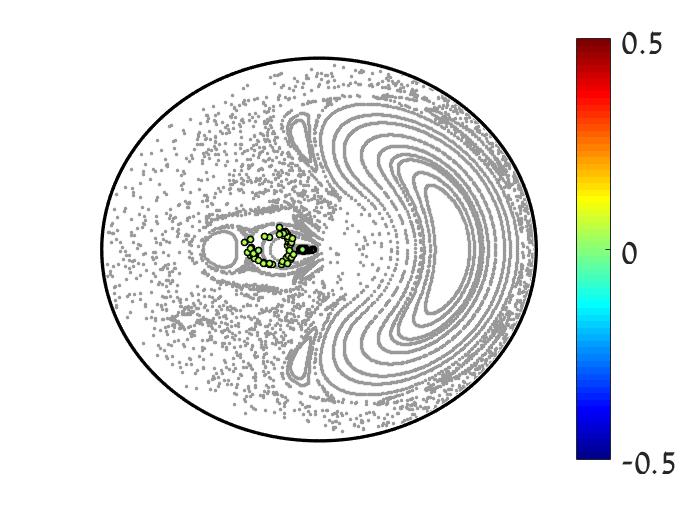}
\\
\includegraphics[width=4cm]{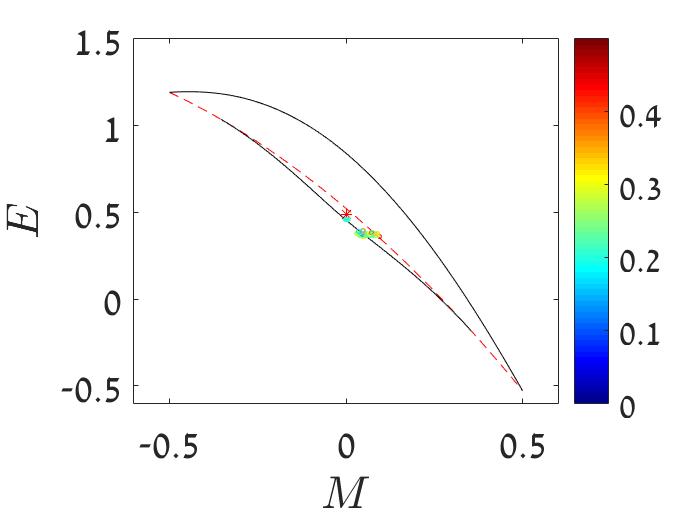}
\includegraphics[width=4cm]{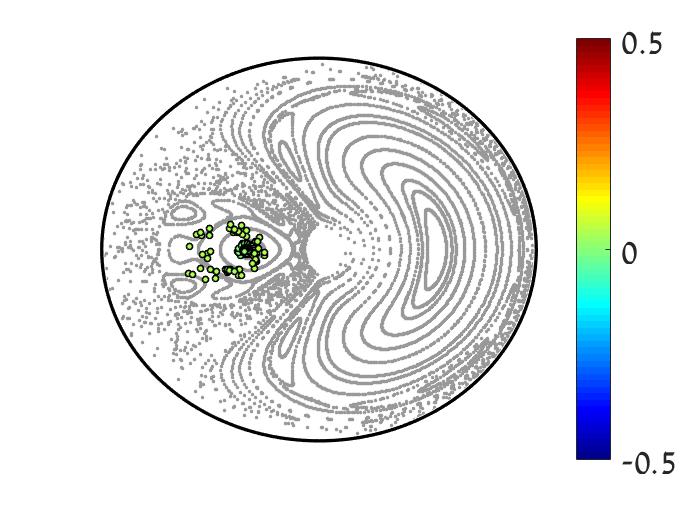}
\\
\includegraphics[width=4cm]{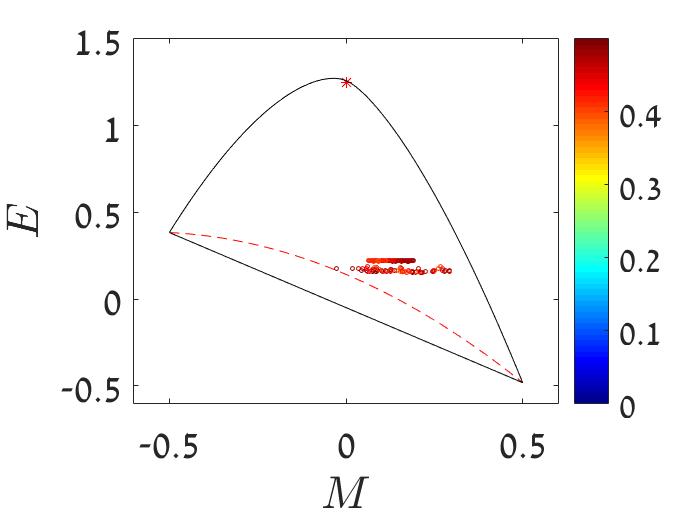}
\includegraphics[width=4cm]{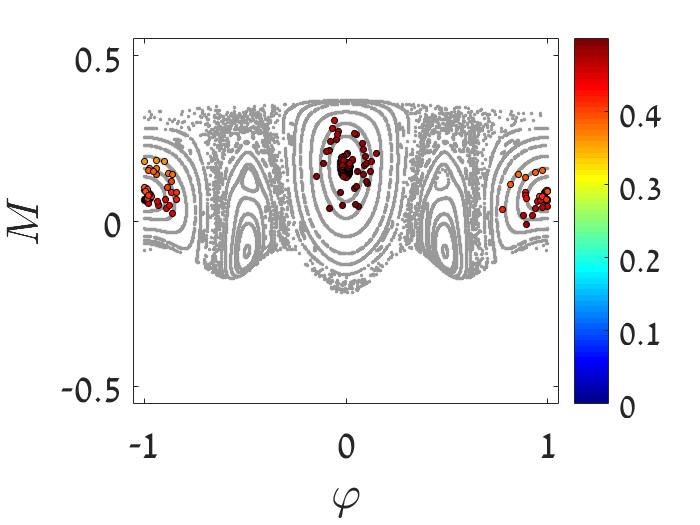}
\\
\includegraphics[width=4cm]{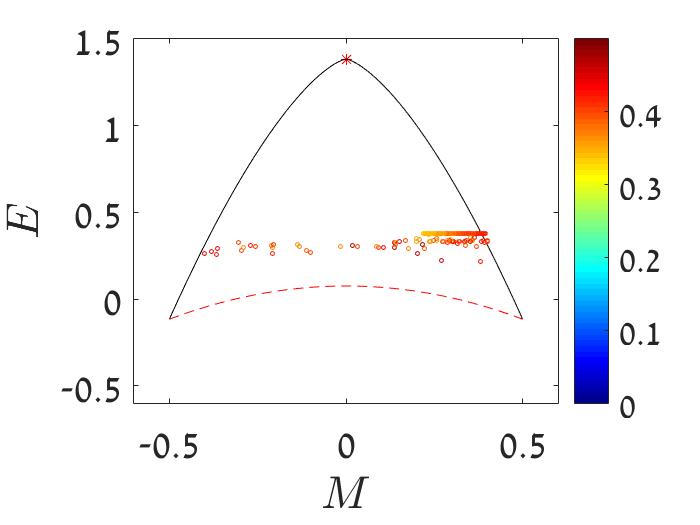}
\includegraphics[width=4cm]{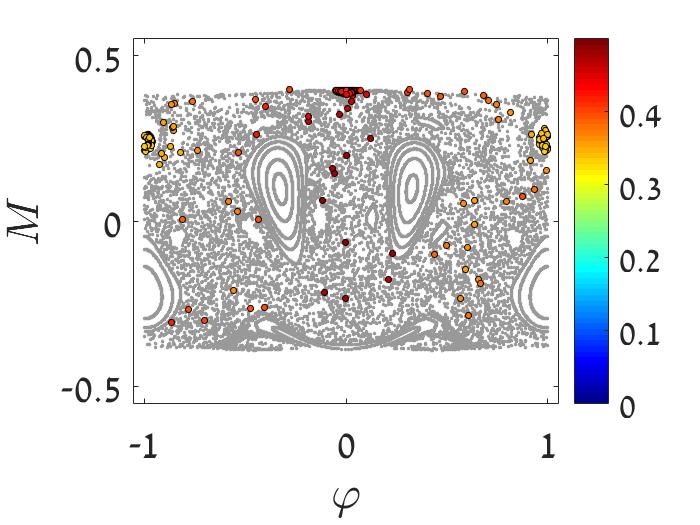}
\\
\includegraphics[width=7cm]{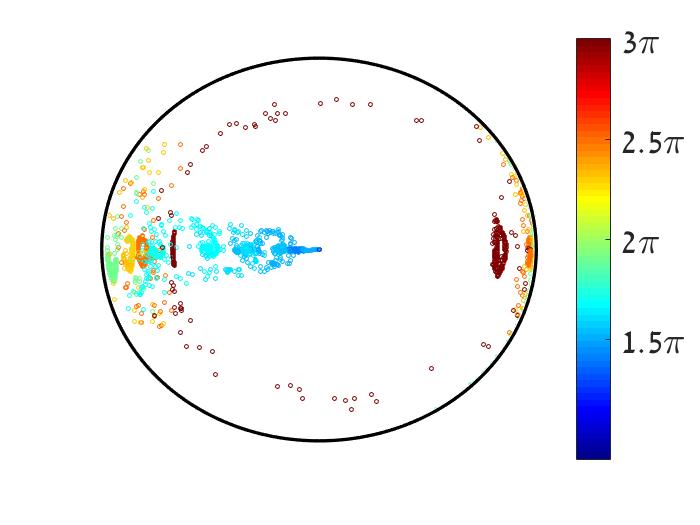}
\end{minipage}
\caption{\label{fig:our}
{\bf Phase space perspective for the simulations of a sweep process.} 
The rates are  $\dot{\Phi}=5\pi \cdot 10^{-4}$ (left set of panels) 
and $\dot{\Phi}=3\pi \cdot 10^{-4}$ (right set of panels).  
The interaction is $u{=}2.3$. 
\rmrk{The initial preparation is a condensate at $n{=}0$ (represented by a red star). 
Initially it is the minimum of the energy landscape (see insets).}
Snapshots are taken after $\Phi_{\text{dyn}}$ is crossed,   
at $\Phi=1.51\pi,1.6\pi,2.5\pi,3\pi$, 
\rmrk{where the central SP in no longer a local minimum, 
and furthermore it is dynamically unstable.    
Consequently the cloud is free to spread away from ${n{=}M{=}0}$.}    
{\em First column of each set:} snapshots of the evolving cloud in $(E,M)$ space, 
where the points are color-coded by~$n$. 
{\em Second column of each set:} The cloud points, color-coded by $M$, 
are overlayed on the $(\varphi,n)$  Poincare section.  
\rmrk{Two panels use non-polar $(\varphi,M)$ coordinates for enhanced resolution}. 
{\em Bottom of each set:} the evolving cloud in $(\varphi,n)$ Poincare coordinates. 
Snapshots of the cloud are taken at different moments, and are color-coded by~$\Phi$. 
Blue is the initial cloud, and red is its final distribution.  
}
\end{figure*}

%%%%%%%%%%%%%%%%%%%%%%%%%%%%%%%%%%%%%%%%%%%%%%%%%%%%%%%%%%%%%%%%%%%%%%%%%%%%%%%%%%%%%%%%%%%%%
\sect{Sweep-related spreading}
We now consider again a quasi-static sweep process. Naively, we might expect that spreading will start once $\Phi_{\text{dyn}}$ is crossed. But a more careful inspections reveals that the QS limit is subtle. We see from the upper left panel of \Fig{fMaEvsTime}, 
and from \Fig{fPS}c that for a slow sweep the cloud splits into two pieces. 
\rmrk{The dynamics is caricatured in \Fig{fDYN}.}  
The reason for the splitting is related to the co-existence of two different mechanisms. 
One resembles the quench scenario. 
Namely, somewhere in the range ${ [\Phi_{\text{stb}}, \Phi_{\text{dyn}}] }$ spreading is initiated along the chaotic strip. 
But a different spreading mechanisms comes into play after $\Phi_{\text{dyn}}$ is crossed. 
This second mechanism dominates the \rmrk{``optimal sweep"} of \Fig{fMaEvsTime}. 
For an optimal sweep the chaos-related spreading mechanism has no time to develop.  

The additional sweep-related mechanism is not related to chaos, 
but to the bifurcation of the stability island. 
It obeys the Kruskal-Neishtadt-Henrard theorem \cite{Kruskal,Neishtadt1,Timofeev,Henrard,Tennyson,Hannay,Cary,Neishtadt2,Elskens,Anglin,Neishtadt3}, 
namely, the cloud is drained into the emerging stability island. 
The full optimal sweep scenario is displayed in the left panels of \Fig{fig:our}.

%%%%%%%%%%%%%%%%%%%%%%%%%%%%%%%%%%%%%%%%%%%%%%%%%%%%%%%%%%%%%%%%%%%%%%%%%%%%%%%%%%%%%%%%%%%%%
\sect{Depletion process}
As we already observed in \Fig{fMaEvsTime}, the spreading of the cloud starts before or latest at $\Phi_{\text{dyn}}$.   
But looking at the color-code we see that the depletion happens at a distinct moment when $\Phi(t) \sim \Phi_{\text{swp}}$. 
This is the moment when a corridor connects the central SP $n{=}0$ with the peripheral 
region $n{=}N/2$. In the absence of chaos $n{=}N/2$ is formally an SP of the ${\mathcal{H}^{(0)}(\varphi,n;M{=}0)}$ Hamiltonian.  
Each SP has its own separatrix. For $\Phi= \Phi_{\text{swp}}$ the two SPs have the same energy, 
and therefore the two separatrices coalesce. From the equation ${ E_0 = E_{\infty}(0) }$ we get 
\beq \label{PhiSWP}
\Phi_{\text{swp}} \ \ = \ \ 3 \arccos\left(-\frac{1}{18}u\right) 
\eeq
Once we add the $\mathcal{H}^{\pm}$  terms,  
this joint separatrix becomes a chaotic strip, what we call ``corridor".
The corridor is available for a small range of $\Phi$ around $\Phi \sim \Phi_{\text{swp}}$. 
During the time interval that the corridor is opened, the central SP is depleted. 
Both the energy landscape and the evolution are demonstrated in \Fig{fig:our}.

%%%%%%%%%%%%%%%%%%%%%%%%%%%%%%%%%%%%%%%%%%%%%%%%%%%%%%%%%%%%%%%%%%%%%%%%%%%%%%%%%%%%%%%%%%%%%
\sect{Subsequent evolution}
We already pointed out that strict classical adiabaticity in the QS sense of Kubo 
does not hold for our scenario: for ${\Phi(t) > \Phi_{\text{swp}}}$ the system does not 
follow any of the $E_n$ curves. The reason for that is figured out by further inspection of the dynamics.
For ${\Phi(t) > \Phi_{\text{swp}}}$ the chaotic strip decomposes 
into quasi regular tori. Consequently a different adiabatic scenario takes over, 
that of Einstein and Landau, where adiabatic invariants are the ``actions" of the tori.
Each piece of the cloud is locked in a different torus, 
and therefore we do not observe in \Fig{fig:our} further ergodization in the $M$~direction.

%%%%%%%%%%%%%%%%%%%%%%%%%%%%%%%%%%%%%%%%%%%%%%%%%%%%%%%%%%%%%%%%%%%%%%%%%%%%%%%%%%%%%%%%%%
\sect{Quasi static average}
Without any approximation we always have ${\dot{E} = -\braket{I}_t \dot{\Phi}}$.
In the Ott-Wilkinson-Kubo formulation of linear response theory \cite{Ott1,Ott2,Ott3,Wilkinson1,Wilkinson2,crs,frc},  
it is assumed that for a QS process the instantaneous average can be replaced by 
an evolving microcanonical average $\braket{I}_{E}$ due to quasi-ergodicity. 
But we are not dealing with a globally chaotic energy surface. 
Rather, the cloud occupies at any moment only a fraction of the energy shall, 
or a set tori that depart from the microcanonical shell. 
We use the notation $\braket{I}_{QS}$ for the corresponding average. Accordingly 
\beq \label{edE}
dE \ \ = - \braket{I}_{QS} \ d\Phi
\eeq
For a system with 2 freedoms the QS average is well defined: 
at any moment the ergodic region that is accessible for the evolving cloud 
is bounded by KAM surfaces. This is not true if we had more than 2 freedoms:
then the accessible region would likely exhibit a more complicated dependence on the rate of the sweep.
Anyway, in the present context the current of \Eq{eq:I} reflects the occupation 
of the orbitals, and therefore can be expressed in terms of ${(M,n)}$. 
The expectation value of the current can be calculated for the evolving cloud of the simulation,  
see inset of \Fig{fMaEvsTime}, and we have verified numerically (not shown) 
that it agrees with \Eq{edE}.

%%%%%%%%%%%%%%%%%%%%%%%%%%%%%%%%%%%%%%%%%%%%%%%%%%%%%%%%%%%%%%%%%%%%%%%%%%%%%%%%%%%%%%%%%%
\sect{Post-sweep ergodization}
For a QS process it is expected to witness quasi-ergodic distribution at any moment. For faster sweep the cloud fails to follow the evolving energy landscape, and therefore a post-sweep ergodization stage is expected, as indeed observed in \Fig{fMaEvsTime} for the ``faster" sweep. But surprisingly post-sweep ergodization stage is also observed if the sweep rate is extremely slow, as observed in \Fig{fMaEvsTime} for the ``slow" sweep. The reason for that is explained by \Fig{fig:our}. Namely, in the case of a very slow dynamics, the cloud is split into several branches as explained previously. Most of it is re-trapped by quasi-integrable tori. But at the very last moment most of the tori are destroyed, and chaos takes-over again. Consequently a fraction of the cloud, that is no longer locked by tori, undergoes post-sweep ergodization.

%%%%%%%%%%%%%%%%%%%%%%%%%%%%%%%%%%%%%%%%%%%%%%%%%%%%%%%%%%%%%%%%%%%%%%%%%%%%%%%%%%%%%%%%%%
%%%%%%%%%%%%%%%%%%%%%%%%%%%%%%%%%%%%%%%%%%%%%%%%%%%%%%%%%%%%%%%%%%%%%%%%%%%%%%%%%%%%%%%%%%
\newpage 
%\vspace*{5mm}
\noindent {\Large\bf\textsl{Discussion}} 
\vspace*{1mm}

%%%%%%%%%%%%%%%%%%%%%%%%%%%%%%%%%%%%%%%%%%%%%%%%%%%%%%%%%%%%%%%%%%%%%%%%%%%%%%%%%%%%%%%%%%%%%
%\sect{Summary}
%
Disregarding the very well studied 2-site Bosonic Josephosn junction, the trimer is possibly the simplest building block for an atomtronic circuit. It is the smallest ring that possibly can be exploited as a SQUID-type Qubit device  \cite{Amico,Paraoanu,Hallwood,sfr}. The first requirement is to have the possibility to witness a stable superflow \cite{sfc,sfa,bhm}. The second requirement is to have the possibility to witness coherent operation. The latter is indicated by, say, coherent oscillations between clockwise and anti-clockwise superflow currents \cite{sfr}. The third requirement is to have the possibility to execute protocols that do not spoil the coherence, meaning that the particles remain condensed in some evolving orbital \cite{cst}. In semiclassical perspective it means that an initial Gaussian cloud does not ergodize.    
One may say that {\em ergodicity} due to chaos, as opposed to {\em stability}, is the threat that looms over the condensation of bosons in optical lattices.

Inspired by experiments with toroidal rings \cite{exprRingNIST}, here we considered a lattice ring that undergoes a prototype sweep protocol: increasing $\Phi$ from~$0$ to~$3\pi$ such that the $k{=}0$ orbital goes from the floor to the ceiling.      
During this process this orbital is depleted. The details of the process are as follows: As $\Phi$ is increased beyond a value $\Phi_{\text{mts}}$, the followed SP becomes a metastable minimum; For $\Phi$ larger that $\Phi_{\text{stb}}$ it becomes a saddle in the energy landscape of the circuit; Depending on the sweep rate it can maintain dynamical stability up to some larger value $\Phi_{\text{dyn}}$; Beyond this value the SP becomes unstable, but this does not automatically implies that the coherent state is depleted; A fully developed depletion process requires a {\em corridor} that leads to ergodization within a chaotic sea; Such corridor is opened  during a small interval around $\Phi \sim \Phi_{\text{swp}}$; During the chaotic stage of the sweep we witness partial ergodization, and the final state of the system is in general not fully-coherent. An optimal sweep rate can be determined.  

In a larger perspective we emphasize that the traditional view of adiabaticity is not enough in order to a address a QSTP for a system that has {\em mixed} integrable and chaotic dynamics. Some historical background is essential in order to appreciate this statement. On the one extreme we have the {\em Einstein-Landau theory} for adiabaticity for integrable systems \cite{Landau}. On the other extreme we have the {\em Kubo-Ott-Wilkinson picture} of adiabaticity in chaotic systems \cite{Ott1,Ott2,Ott3,Wilkinson1,Wilkinson2,crs,frc}, which is associated with energy absorption in accordance with linear-response theory. But realistic systems are neither integrable nor chaotic, but rather have mixed phase space whose topological structure changes during the sweep process. The simplest scenario is separatrix crossing, that can be addressed using the Kruskal-Neishtadt-Henrard theorem \cite{Kruskal,Neishtadt1,Timofeev,Henrard,Tennyson,Hannay,Cary,Neishtadt2,Elskens,Anglin,Neishtadt3}. More generally tori can merge into chaos, and new sets of tori can be formed later on.  This leads to anomalous dissipation \cite{Kedar1,Kedar2} and irreversibility in the QS limit \cite{apc,lbt}. With the same spirit  we have explored in this work the mechanisms that are involved in QS {\em transfer} protocols, and also the non-trivial dependence of the outcome on the sweep rate.

%%%%%%%%%%%%%%%%%%%%%%%%%%%%%%%%%%%%%%%%%%%%%%%%%%%%%%%%%%%%%%%%%%%%%%%%%%%%%%%%%%%%%%%%%%
%%%%%%%%%%%%%%%%%%%%%%%%%%%%%%%%%%%%%%%%%%%%%%%%%%%%%%%%%%%%%%%%%%%%%%%%%%%%%%%%%%%%%%%%%%
\vspace*{5mm}
%\newpage
\noindent {\Large\bf\textsl{Methods}} 
\vspace*{1mm}

%%%%%%%%%%%%%%%%%%%%%%%%%%%
\sect{The Hamiltonian} 
The BHH for an $L$-site rotating ring is 
\beq
\mathcal{H} = \sum_{j=1}^{L} \left[
\frac{U}{2}  {a}_{j}^{\dag}  {a}_{j}^{\dag}  {a}_{j} {a}_{j} 
- \frac{K}{2} \left(\eexp{i(\Phi/L)}  {a}_{j{+}1}^{\dag} {a}_{j} + \text{h.c.} \right)
\right] \ \ \ 
\eeq
where $j$ mod$(L)$ labels the sites of the ring, 
the $a$-s are the bosonic field operators, 
and $\Phi$ is the Sagnac phase.

It is convenient to switch to momentum representation. 
For a clean ring the momentum orbitals 
have wavenumbers $k=(2\pi/L)\times \text{integer}$. 
One defines annihilation and creation operators ${b}_{k}$ and ${b}_{k}^{\dag}$,  
such that $b_k^{\dag} = \frac{1}{ \sqrt{L} } \sum_j \eexp{ikj} a_j^{\dag}$ 
creates bosons in the $k$-th momentum orbitals.
Consequently the BHH takes the form 
\beq
\mathcal{H} \ \ = \ \ \sum_{k} \epsilon_k b_k^{\dag}b_k \ + \ \frac{U}{2L} \sum' b_{k_4}^{\dag}b_{k_3}^{\dag}b_{k_2}b_{k_1}
\eeq
where the constraint ${k_1{+}k_2{+}k_3{+}k_4=0}$ mod($2\pi$) is indicate  
by the prime, and the single particle energies are
\beq
\epsilon_k \ = \ -K \cos\left(k- \frac{\Phi}{L} \right)
\eeq 
Later we assume, without loss of generality, that the particles are initially condensed 
in the ${k=0}$ orbital. This is not necessarily the ground-state orbital, 
because we keep $\Phi$ as a free parameter. 
Note that we optionally use~$k$ as a dummy index to label the momentum orbitals.

%%%%%%%%%%%%%%%%%%%%%%%%%%%
\sect{Trimer Hamiltonian} 
For the purpose of semiclassical treatment we express the Hamiltonian in terms of occupations and conjugate phases.  
For the  3-site ring ($L{=}3$) we get: 
\beq %\nonumber
\mathcal{H} &=& 
\sum_{k=0,1,2} \epsilon_k n_k \ + \ \frac{U}{6} \sum_k n_k^2 \ + \  \frac{U}{3} \sum_{k'\ne k} n_{k'} n_{k} 
\\ \label{eA4} \nonumber
&+& \frac{U}{3} \sum_{k'' \ne k' \ne k}  \left[n_{k'}n_{k''}\right]^{1/2}  \ n_{k}  \ \cos \left( \varphi_{k''} + \varphi_{k'} - 2 \varphi_{k} \right)  
\ \ \ \
\eeq
We define ${q_1 = \varphi_1 - \varphi_0 }$ and ${q_2 = \varphi_2 - \varphi_0 }$
where the subscripts refers to ${k_{1,2}=\pm(2\pi/3)}$.  
Using the notation 
\beq
\mathcal{E}_k \ = \ (\epsilon_{k}-\epsilon_{0}) + (1/3)NU
\eeq
we get ${\mathcal{H} = \mathcal{H}^{(0)} +  \left[\mathcal{H}^{(+)} + \mathcal{H}^{(-)} \right]}$ with
\beq \nonumber
\mathcal{H}^{(0)} &=& 
\  \epsilon_{0}N + \frac{U}{6}N^2 
\ + \ \mathcal{E}_1 n_1 + \mathcal{E}_2 n_2 
\\ \nonumber &-& \ \frac{U}{3} \left[ n_1^2 + n_2^2 + n_1n_2 \right]  
\\ \label{eHo} &+& \ \frac{2U}{3} (N{-}n_1{-}n_2) \sqrt{n_1 n_2}  \cos\left(q_1 + q_2\right) 
\ \ \ \
\eeq
and
\beq
\mathcal{H}^{(+)} = \frac{2U}{3} \sqrt{(N{-}n_1{-}n_2) n_1} \ n_2 \cos \left(q_1 - 2 q_2 \right)
\ \ \ \ 
\eeq
while $\mathcal{H}^{(-)}$ is obtained by swapping the indices~(${1 \leftrightarrow 2}$).

%%%%%%%%%%%%%%%%%%%%%%%%%%%%%%%%%%
\sect{Compact form} 
It is more convenient to use the coordinates  
\beq \nonumber
\phi[\text{mod}(4\pi)] & \ = \ & q_1-q_2  \ =  \  \varphi_1 - \varphi_2 \\
\varphi[\text{mod}(2\pi)] & \ = \ & q_1+q_2 \ =  \ \varphi_1 + \varphi_2 - 2 \varphi_0 
\eeq
and the conjugate coordinates 
\beq
M & \ = \ & \frac{1}{2}(n_{1} - n_{2}) \ \ \in \left[-\frac{N}{2},\frac{N}{2}\right] \\
n & \ = \ & \frac{1}{2}(n_{1} + n_{2}) \ \ \in \left[|M|, \frac{N}{2}\right] 
\eeq
Then the Hamiltonian takes the form of \Eq{eHfull}
with \Eq{eH0} and \Eq{eHchaos}. 
The energy $E_0$ of the $n{=}0$ central SP is implied by the first two terms of \Eq{eHo}, 
leading to \Eq{eE0}. The detuning parameters are 
\beq
\mathcal{E}_{\parallel} &=& \mathcal{E}_1+\mathcal{E}_2 - (1/2)NU \\ 
\mathcal{E}_{\perp} &=& \mathcal{E}_1 - \mathcal{E}_2  
\eeq
leading to \Eq{eEDn} and \Eq{eEDM}.
Note: if we linearized $\mathcal{H}$ with respect to the ${(n_1,n_2)}$ occupations, 
we would get the Bogolyubov approximation, which is \Eq{eH0}
without the third term ($M^2$), and with ${(N{-}2n)\approx N}$.

\begin{figure}[t!]
\centering 
\includegraphics[width=7cm]{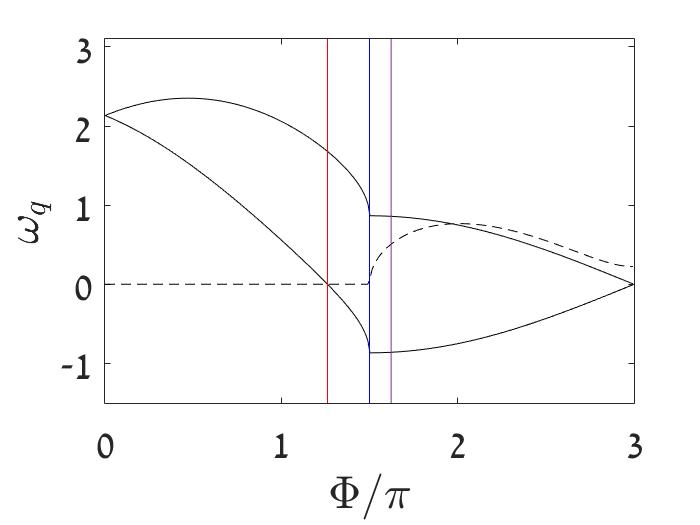} 
\caption{ 
{\bf The Bogolyubov frequencies.} 
They are calculated for a ${k{=}0}$ condensate.
The vertical lines from left to right are for $\Phi_{\text{stb}}$, 
and $\Phi_{\text{dyn}}$, and $\Phi_{\text{swp}}$. 
The latter cannot be deduced form the Bogolyubov analysis, 
but requires global understanding of phase space structure.  
}	
\label{fBogo}
\end{figure}

%%%%%%%%%%%%%%%%%%%%%%%%%%%%%%%%%%%%%%%%%%%%%%%%%%%%%%%%%%%%
\sect{Bogolyubov frequencies} 
The non-trivial Bogolyubov frequencies in units of $K=1$, see \rmrk{SM}, are  
\beq \label{eBog} 
\omega_{\pm} = \pm \frac{\sqrt{3}}{2}\sin{\frac{\Phi}{3}} + \sqrt{\left(\frac{3}{2}\cos{\frac{\Phi}{3}}\right)^2+u\cos{\frac{\Phi}{3}}}
\ \ \ \ \ \ \ 
\eeq
For positive $u$ and $\Phi<\Phi_{\text{stb}}$ the SP is the minimum 
of the energy landscape, and the Bogolyubov frequencies are positive, see \Fig{fBogo}.  
The SP becomes a saddle once $\omega_{-}$ changes sign and becomes negative. 
The SP becomes dynamically unstable once the $\omega_{\pm}$ become complex. 
Note that the energy of the SP, once it becomes unstable, 
gets above the $M{=}0$ floor, see \Fig{fLandscape} of \rmrk{SM}.
By inspection of \Eq{eBog} we can identify a critical value of the interaction ${u_c=9/4}$.
For large interaction (${u>u_c}$) the Bogolyubov frequencies remain complex up to the 
end of the sweep at ${\Phi=3\pi}$. This indicates that the SP in not at the maximum of 
the energy landscape, see \Fig{fLandscape}. The upper most SPs in this region support self-trapped states.  
For weak interaction (${u<u_c}$) the Bogolyubov frequencies become real 
and negative once we cross ${\Phi=3\arccos{\left(-(9/4)u\right)}}$, 
indicating that the SP becomes a stable maximum.

%%%%%%%%%%%%%%%%%%%%%%%%%%%%%%%%%%%
% Energy levels for u=1.0, 2.3, 4.5
% Bogo for the same plot
% Quantum simulation
% Current I vs time

%%%%%%%%%%%%%%%%%%%%%%%%%%%%%%%%%%%%%%%%%%%%%%%%%%%%%
\clearpage
%\newpage

%%%%%%%%%%%%%%%%%%%%%%%%%%%%%%%
\sect{Acknowledgment} 
This research was supported by the Israel Science Foundation (Grant  No.283/18).
D.C acknowledge the cooperation with Amichay Vardi on related problems. 
\\

\sect{Contributions}
Both authors have contributed to this article. Y.W has carried out the analysis, including numerics and figure preparation. The themes of the study and the text of the Ms have been discussed, written and iterated jointly by D.C. and Y.W.
\\

\sect{Competing interests}
The authors declare no competing financial interests.
\\

\sect{Corresponding authors} 
Correspondence to D.C. [dcohen@bgu.ac.il].

% %%% to reset bib numbers:
% \setcounter{enumiv}{0}
% \setcounter{NAT@ctr}{0}

%\bibliographystyle{apsrev4-1}
%\bibliography{mybib}

\begin{thebibliography}{99}

%%% adiabticity, regular phase space 

% adiabatic theory - Landau
\bibitem{Landau} 
L.D.~Landau, E.M.~Lifshitz, {\em Mechanics}, 3rd.~Ed., p.~154ff. Elsevier (1982).


%%% adiabticity, chaotic phase space 
%\cite{Ott1,Ott2,Ott3,Wilkinson1,Wilkinson2,crs,frc}.

% adiabatic theory - Ott
\bibitem{Ott1} 
E.~Ott, 
{\em Goodness of ergodic adiabatic invariants},
Phys. Rev. Lett. {42}, 1628 
\hrefl{1979}{https://doi.org/10.1103/PhysRevLett.42.1628}

% adiabatic theory - Ott
\bibitem{Ott2} 
R.~Brown, E.~Ott, C.~Grebogi, 
{\em Ergodic adiabatic invariants of chaotic systems},
Phys. Rev. Lett, {59}, 1173 
\hrefl{1987}{https://doi.org/10.1103/PhysRevLett.59.1173}

% adiabatic theory - Ott
\bibitem{Ott3} 
R.~Brown, E.~Ott, C.~Grebogi, 
{\em The goodness of ergodic adiabatic invariants}
J. Stat. Phys. {49}, 511 
\hrefl{1987}{https://doi.org/10.1007/BF01009347}

% LRT - Wilkinson
\bibitem{Wilkinson1} 
M.~Wilkinson, 
{\em A semiclassical sum rule for matrix elements of
classically chaotic systems}, J. Phys. A {20}, 2415 
\hrefl{1987}{https://doi.org/10.1088/0305-4470/20/9/028}

% LRT - Wilkinson
\bibitem{Wilkinson2} 
M.~Wilkinson, 
{\em Statistical aspects of dissipation by Landau-Zener transitions}, J. Phys. A {21}, 4021 
\hrefl{1988}{https://doi.org/10.1088/0305-4470/21/21/011}


%crs
\bibitem{crs}
D. Cohen, 
{\em Quantum Dissipation due to the interaction with chaotic degrees-of-freedom and the correspondence principle},
Phys. Rev. Lett. {82}, 4951 
\hrefl{1999}{http://dx.doi.org/10.1103/PhysRevLett.82.4951}

%frc
\bibitem{frc}          
D. Cohen, 
{\em Chaos and Energy Spreading for Time-Dependent Hamiltonians, and the various Regimes in the Theory of Quantum Dissipation},
Annals of Physics {283}, 175-231 
\hrefl{2000}{http://dx.doi.org/10.1006/aphy.2000.6052} 




%%%% adiabticity,  mixed phase space 
% \cite{Kedar1,Kedar2}
% \cite{apc,lbt}

%\bibitem
%K. Shah, V. Gelfreich, V. Rom-Kedar, D. Turaev,  
%Leaky Fermi Accelerators,
%Phys. Rev. E 91, 062920, 
%\hrefl{2015}{https://journals.aps.org/pre/abstract/10.1103/PhysRevE.91.062920}

% adiabatic theory - Kedar
\bibitem{Kedar1}
V.~Gelfreich, V.~Rom-Kedar, D.~Turaev,  
{\em Oscillating mushrooms: adiabatic theory for a non-ergodic system}, 
JJ. Phys. A {47}, 395101
\hrefl{2015}{https://doi.org/10.1088/1751-8113/47/39/395101}

% fast system is chaotic
\bibitem{Kedar2}
K.~Shah, D.~Turaev, V.~Gelfreich, V.~Rom-Kedar, 
{\em Equilibration of energy in slow-fast systems}, 
PNAS {114(49)}, E10514, 
\hrefl{2017}{https://doi.org/10.1073/pnas.1706341114}

%apc
\bibitem{apc} 
A. Dey, D. Cohen, A. Vardi,
{\em Adiabatic passage through chaos}, 
Phys. Rev. Lett. {121}, 250405 
\hrefl{2018}{https://journals.aps.org/prl/abstract/10.1103/PhysRevLett.121.250405}
%DOI:https://doi.org/10.1103/PhysRevLett.121.250405 

%lbt
\bibitem{lbt} 
R. Burkle, A. Vardi, D. Cohen, J.R. Anglin, 
{\em Probabilistic hysteresis in isolated integrable and chaotic Hamiltonian systems}, 
Phys. Rev. Lett. {123}, 114101 
\hrefl{2019}{https://journals.aps.org/prl/abstract/10.1103/PhysRevLett.123.114101}
%http://arxiv.org/abs/1904.00474
%DOI:https://doi.org/10.1103/PhysRevLett.123.114101



%%%% adiabticity, spepratrix crossing 
%\cite{Kruskal,Neishtadt1,Timofeev,Henrard,Tennyson,Hannay,Cary,Neishtadt2,Elskens,Anglin,Neishtadt3}.

% Kruskal
\bibitem{Kruskal} 
D.~Dobbrott, J. M.~Greene, 
{\em Probability of Trapping-State Transition in a Toroidal Device},
Phys. of Fluids {14}, 7 
\hrefl{1971}{https://doi.org/10.1063/1.1693639}.

% separatrix crossing
\bibitem{Neishtadt1} A. I.~Neishtadt, 
{\em Passage through a separatrix in a resonance problem with a slowly-varying parameter}, 
J. Appl. Math. Mech. {39}, 594-605 
\hrefl{1975}{https://doi.org/10.1016/0021-8928(75)90060-X}.

\bibitem{Timofeev} 
A.V.~Timofeev, 
{\em On the constancy of an adiabatic invariant when the nature of the motion changes}, 
JETP {48}, 656 \hrefl{1978}{http://www.jetp.ac.ru/cgi-bin/dn/e_048_04_0656.pdf}.

\bibitem{Henrard} 
J.~Henrard, 
{\em Capture into resonance: an extension of the use of adiabatic invariants}, Celestial Mechanics {27}, 3-22 
\hrefl{1982}{https://doi.org/10.1007/BF01228946}.

\bibitem{Tennyson} 
J.R.~Cary, J.~R., D.F.~Escande, J.L.~Tennyson,
{\em Adiabatic-invariant change due to separatrix crossing}, 
Phys. Rev. A {34}, 4256--4275 \hrefl{1986}{https://doi.org/10.1103/PhysRevA.34.4256}.

\bibitem{Hannay} 
J.H~Hannay, {\em Accuracy loss of action invariance in adiabatic change of a one-freedom Hamiltonian}, 
J. Phys. A {19}, L1067--L1072 \hrefl{1986}{https://doi.org/10.1088/0305-4470/19/17/004}.

\bibitem{Cary} 
J.R.~Cary, R.T.~Skodje, 
{\em Reaction probability for sequential separatrix crossings}, 
Phys. Rev. Lett. {61}, 1795--1798 \hrefl{1991}{https://doi.org/10.1103/PhysRevLett.61.1795}.

\bibitem{Neishtadt2} 
A.I.~Neishtadt, 
{\em Probability phenomena due to separatrix crossing}, 
Chaos {1}, 42 
\hrefl{1991}{https://doi.org/10.1063/1.165816}.

\bibitem{Elskens} 
Y.~Elskens, D.F.~Escande, 
{\em Slowly pulsating separatrices sweep homoclinic tangles where islands must be small: an extension of classical adiabatic theory}, 
Nonlinearity {4}, 615--667 \hrefl{1991}{https://doi.org/10.1088/0951-7715/4/3/002}.

\bibitem{Anglin}
T.~Eichmann, E.P.~Thesing, J.R.~Anglin,   
{\em Engineering separatrix volume as a control technique for dynamical
transitions}. Phys. Rev. E {98}, 052216  
\hrefl{2018}{https://doi.org/10.1103/PhysRevE.98.052216}

\bibitem{Neishtadt3} 
A.~Neishtadt, {\em On mechanisms of destruction of adiabatic invariance in slow–fast Hamiltonian systems}, Nonlinearity {32} (11), R53
\hrefl{2019}{https://doi.org/10.1088/1361-6544/ab2a2c}.


%%% BHH - experiments

%\cite{Oberthaler,Steinhauer}
%\cite{exprBHH1,exprBHH}

\bibitem{Oberthaler}
M.~Albiez, R.~Gati, J.~Folling, S.~Hunsmann, M.~Cristiani, M. K.~Oberthaler, 
{\em Direct Observation of Tunneling and Non-linear Self-Trapping in a Single Bosonic Josephson Junction},
Phys. Rev. Lett. {95}, 010402  
\hrefl{2005}{https://doi.org/10.1103/PhysRevLett.95.010402}.

\bibitem{Steinhauer}
S.~Levy, E.~Lahoud, I.~Shomroni, J.~Steinhauer, {\em The ac
and dc josephson effects in a bose–einstein condensate}, Nature
(London) {449}, 579 
\hrefl{2007}{https://doi.org/10.1038/nature06186}.

\bibitem{exprBHH1}
O.~Morsch, M.~Oberthaler, {\em Dynamics of Bose-Einstein
condensates in optical lattices}, Rev. Mod. Phys. {78}, 179
\hrefl{2006}{https://doi.org/10.1103/RevModPhys.78.179}.

\bibitem{exprBHH2}
I.~Bloch, J.~Dalibard, W.~Zwerger, {\em Many-body
physics with ultracold gases}, Rev. Mod. Phys. {80}, 885
\hrefl{2008}{https://doi.org/10.1103/RevModPhys.80.885}.

%csd
\bibitem{csd} 
M. Chuchem, K. Smith-Mannschott, M. Hiller, T. Kottos, A. Vardi, D. Cohen,
{\em Quantum dynamics in the bosonic Josephson junction},
Phys. Rev. A {82}, 053617 
\hrefl{2010}{http://dx.doi.org/10.1103/PhysRevA.82.053617}


%%% BHH - Ring

% \cite{Amico,sfa,bhm,sfr,Paraoanu,Hallwood,gallemi}

\bibitem{Amico}
L.~Amico, D.~Aghamalyan, F.~Auksztol, H.~Crepaz, R.~Dumke, L. C.~Kwek, 
{\em Superfluid qubit systems with ring shaped optical lattices}, Sci. Rep. {4}, 04298 
\hrefl{2014}{https://doi.org/10.1038/srep04298}.

\bibitem{Paraoanu}
Gh.-S.~Paraoanu, 
{\em Persistent currents in a circular array of bose-einstein condensates}, 
Phys. Rev. A {67}, 023607 
\hrefl{2003}{https://doi.org/10.1103/PhysRevA.67.023607}.

\bibitem{Hallwood}
D. W.~Hallwood, K.~Burnett, J.~Dunningham, 
{\em Macroscopic superpositions of superfluid flows}, 
New J. Phys. {8}, 180 
\hrefl{2006}{https://doi.org/10.1088/1367-2630/8/9/180}.

% sfr 
\bibitem{sfr} 
G. Arwas, D. Cohen,
{\em Chaos and two-level dynamics of the Atomtronic Quantum Interference Device},
New J. Phys. {18}, 015007 
\hrefl{2016}{http://dx.doi.org/10.1088/1367-2630/18/1/015007}





%%%%%%%%%%%%%%%%%%%%%%%%%
%%% BHH - Ring - 3sites


\bibitem{ref12}  % DNLS 
%J.C. Eilbeck, P.S. Lomdahl, A.C. Scott,
Eilbeck, J. C. Lomdahl, P. S. \& Scott, A. C.
{\em The discrete self-trapping equation},
{Physica D} {16} 318-38 
\hrefl{1985}{http://www.sciencedirect.com/science/article/pii/0167278985900120}

%\bibitem{trimer1} %NLS
%J. C. Eilbeck, G. P. Tsironis, and S. K. Turitsyn,
%{\em Stationary states in a doubly nonlinear trimer model of optical couplers},
%Phys. Scr. {52}, 386 (1995).

\bibitem{trimer2} %NLS
%D. Hennig, H. Gabriel, M.F. Jorgensen, P.L. Christiansen, C.B. Clausen,
Hennig, D., Gabriel, H., Jorgensen, M. F., Christiansen, P. L. \& Clausen, C. B.
{\em Homoclinic chaos in the discrete self-trapping trimer},
{Phys. Rev. E} {51}, 2870 
\hrefl{1995}{http://journals.aps.org/pre/abstract/10.1103/PhysRevE.51.2870}

\bibitem{trimer3}
%S. Flach and V. Fleurov,
Flach, S. \& Fleurov, V.
{\em Tunnelling in the nonintegrable trimer - a step towards quantum breathers},
{J. Phys.: Condens. Matter} {9}, 7039 
\hrefl{1997}{http://iopscience.iop.org/0953-8984/9/33/007}

\bibitem{trimer4}
%K. Nemoto, C.A. Holmes, G.J. Milburn, and W.J. Munro,
Nemoto, K., Holmes, C. A., Milburn, G. J. \& Munro, W. J.
{\em Quantum dynamics of three coupled atomic Bose-Einstein condensates},
{Phys. Rev. A} {63}, 013604 
\hrefl{2000}{http://journals.aps.org/pra/abstract/10.1103/PhysRevA.63.013604}

%\bibitem{trimer5}
%R. Franzosi, V. Penna,
%{\em Self-trapping mechanisms in the dynamics of three coupled Bose-Einstein condensates},
%Phys. Rev. A 65, 013601 (2002).
%\hrefl{http://journals.aps.org/pra/abstract/10.1103/PhysRevA.65.013601}


\bibitem{trimer6} %NLS
%R. Franzosi, V. Penna,
Franzosi, R. \& Penna, V.
{\em Chaotic behavior, collective modes, and self-trapping in the dynamics of three coupled Bose-Einstein condensates},
{Phys. Rev. E} {67}, 046227 
\hrefl{2003}{http://journals.aps.org/pre/abstract/10.1103/PhysRevE.67.046227}

\bibitem{trimer15}
%M. Johansson,
Johansson, M.
{\em Hamiltonian Hopf bifurcations in the discrete nonlinear Schrödinger trimer: oscillatory instabilities, quasi-periodic solutions and a new type of self-trapping transition},
{J. Phys. A: Math. Gen.} {37}, 2201-2222 
\hrefl{2004}{http://iopscience.iop.org/0305-4470/37/6/017}


\bibitem{trimer7}
%M. Hiller, T. Kottos, and T. Geisel,
Hiller, M., Kottos, T. \& Geisel, T.
{\em Complexity in parametric Bose-Hubbard Hamiltonians and structural analysis of eigenstates},
{Phys. Rev. A} {73}, 061604(R) 
\hrefl{2006}{http://journals.aps.org/pra/abstract/10.1103/PhysRevA.73.061604}

\bibitem{trimer19}
%C. Lee, T. J. Alexander, and Y.S. Kivshar,
Lee, C., Alexander, T. J. \& Kivshar, Y. S.
{\em Melting of Discrete Vortices via Quantum Fluctuations},
{Phys. Rev. Lett.} {97}, 180408 
\hrefl{2006}{http://journals.aps.org/prl/abstract/10.1103/PhysRevLett.97.180408}

%\bibitem{trimer8}
%E. M. Graefe, H. J. Korsch, and D. Witthaut,
%{\em Mean-field dynamics of a Bose-Einstein condensate in a time-dependent triple-well trap: Nonlinear eigenstates, Landau-Zener models, and stimulated Raman adiabatic passage}
%Phys. Rev. A {73}, 013617 (2006).
%\hrefl{http://journals.aps.org/pra/abstract/10.1103/PhysRevA.73.013617}

%\bibitem{trimer17}
%S. Mossmann and C. Jung,
%Semiclassical approach to Bose-Einstein condensates in a triple well potential
%Phys. Rev. A 74, 033601 (2006)
%\hrefl{http://journals.aps.org/pra/abstract/10.1103/PhysRevA.74.033601}


\bibitem{trimerSREP1}
Chaohong Lee, Tristram J. Alexander, and Yuri S. Kivshar,
{\em Melting of Discrete Vortices via Quantum Fluctuations},
Phys. Rev. Lett. 97, 180408 
\hrefl{2006}{https://journals.aps.org/prl/abstract/10.1103/PhysRevLett.97.180408}


\bibitem{trimer20}
%A.R. Kolovsky,
Kolovsky, A. R.
{\em Semiclassical Quantization of the Bogoliubov Spectrum},
{Phys. Rev. Lett.} {99}, 020401 
\hrefl{2007}{http://journals.aps.org/prl/abstract/10.1103/PhysRevLett.99.020401}

%\bibitem{trimer9}
%J. D. Bodyfelt, M. Hiller, and T. Kottos,
%{\em Engineering fidelity echoes in Bose-Hubbard Hamiltonians},
%Europhys. Lett. {78}, 50003 (2007).
%\hrefl{http://iopscience.iop.org/0295-5075/78/5/50003}

%\bibitem{trimer10}
%P. Buonsante, V. Penna,
% Some remarks on the coherent-state variational approach to nonlinear boson models
%% (derivation of trimer classical equations from the Heisenberg ones, section on Chaoticity)
%J. Phys. A 41, 175301 (2008)
%\hrefl{http://iopscience.iop.org/1751-8121/41/17/175301}

%\bibitem{trimer11} 
%M. Hiller, T. Kottos, and T. Geisel, 
%{\em Wave-packet dynamics in energy space of a chaotic trimeric Bose-Hubbard system}
%Phys. Rev. A {79}, 023621 (2009).
%\hrefl{http://journals.aps.org/pra/abstract/10.1103/PhysRevA.79.023621}


%\bibitem{trimer16}
%F. Trimborn, D. Witthaut, and H. J. Korsch,
%Beyond mean-field dynamics of small Bose-Hubbard systems based on the number-conserving phase-space approach
%Phys. Rev. A 79, 013608 (2009)
%\hrefl{http://journals.aps.org/pra/abstract/10.1103/PhysRevA.79.013608}

\bibitem{trimer18}
%P. Buonsante, V. Penna, and A. Vezzani,
Buonsante, P., Penna, V. \& Vezzani, A.,
{\em Quantum signatures of the self-trapping transition in attractive lattice bosons},
{Phys. Rev. A} {82}, 043615 
\hrefl{2010}{http://journals.aps.org/pra/abstract/10.1103/PhysRevA.82.043615}


\bibitem{trimer12}
%T.F. Viscondi, K. Furuya,
Viscondi, T. F. \& Furuya, K.
{\em Dynamics of a Bose–Einstein condensate in a symmetric triple-well trap},
J. Phys. A {44}, 175301 
\hrefl{2011}{http://iopscience.iop.org/1751-8121/44/17/175301}

%\bibitem{trimer14}
%A. P. Itin and P. Schmelcher
%Semiclassical spectrum of small Bose-Hubbard chains: A normal-form approach
%Phys. Rev. A 84, 063609 (2011)
%\hrefl{http://journals.aps.org/pra/abstract/10.1103/PhysRevA.84.063609}

\bibitem{trimer13}
%P. Jason, M. Johansson, K. Kirr,
Jason, P., Johansson, M. \& Kirr, K.
{\em Quantum signatures of an oscillatory instability in the Bose-Hubbard trimer},
Phys. Rev. E 86, 016214 
\hrefl{2012}{http://link.aps.org/doi/10.1103/PhysRevE.86.016214}


\bibitem{trimerSREP2}
L. Morales-Molina, S.A. Reyes, and M. Orszag, 
{\em Current and entanglement in a three-site Bose-Hubbard ring},
Phys. Rev. A 86, 033629 
\hrefl{2012}{https://journals.aps.org/pra/abstract/10.1103/PhysRevA.86.033629}


\bibitem{gallemi} 
A.~Gallemí, M.~Guilleumas, J.~Martorell, R.~Mayol, A.~Polls, B.~Juliá-Díaz, 
{\em Fragmented condensation in Bose–Hubbard trimers with tunable tunnelling}, 
New J. Phys. {17}, 073014
\hrefl{2015}{https://doi.org/10.1088/1367-2630/17/7/073014}.


%%% trimer (ours)

\bibitem{sfs}
%G. Arwas, A. Vardi, D. Cohen, 
Arwas, G., Vardi, A. \& Cohen, D.
Triangular Bose-Hubbard trimer as a minimal model for a superfluid circuit,
{Phys. Rev. A} {89}, 013601 
\hrefl{2014}{http://journals.aps.org/pra/abstract/10.1103/PhysRevA.89.013601}

%sfc
\bibitem{sfc} 
G. Arwas, A. Vardi, D. Cohen,
{\em Superfluidity and Chaos in low dimensional circuits},
Scientific Reports {5}, 13433 
\hrefl{2015}{http://dx.doi.org/10.1038/srep13433}


% sfa 
\bibitem{sfa}
G. Arwas, D. Cohen,
{\em Superfluidity in Bose-Hubbard circuits},
Phys. Rev. B {95}, 054505  
\hrefl{2017}{http://link.aps.org/doi/10.1103/PhysRevB.95.054505}
%DOI: 10.1103/PhysRevB.95.054505
%https://arxiv.org/abs/1612.00251

%bhm
\bibitem{bhm} 
G. Arwas, D. Cohen, 
{\em Monodromy and chaos for condensed bosons in optical lattices},
Phys. Rev. A {99}, 023625  
\hrefl{2019}{https://journals.aps.org/pra/abstract/10.1103/PhysRevA.99.023625}
%DOI:https://doi.org/10.1103/PhysRevA.99.023625





%%%%%%%%%%%%%%%%%%%%%%%%%%%%%%%%%
%%  underlying mixed phase space

%%% \cite{KolovskyReview,sfc}

\bibitem{KolovskyReview}
A. R.~Kolovsky, 
{\em Bose-Hubbard hamiltonian: Quantum chaos
approach}, Int. J. Mod. Phys. B {30}, 1630009 
\hrefl{2016}{https://doi.org/10.1142/S0217979216300097}.



%%%%%%%%%%%%%%%%%%%%%%%%%%%%%%%%%
%%  sweep / swallow

%%% \cite{Swallow1,Swallow2,Swallow3,Swallow4,Swallow5}

\bibitem{Swallow1}
E.J. Mueller,
{\em Superfluidity and mean-field energy loops: Hysteretic behavior in Bose-Einstein condensates},
Phys. Rev. A 66, 063603
\hrefl{2002}{https://journals.aps.org/pra/abstract/10.1103/PhysRevA.66.063603}

\bibitem{Swallow2}
B. Wu and Q. Niu, 
{\em Superfluidity of Bose–Einstein condensate in an optical lattice: Landau–Zener tunnelling and dynamical instability},
New. J. Phys. 5, 104 
\hrefl{2003}{https://iopscience.iop.org/article/10.1088/1367-2630/5/1/104}

\bibitem{Swallow3}
M. Machholm, C.J. Pethick, H. Smith,
{\em Band structure, elementary excitations, and stability of a Bose-Einstein condensate in a periodic potential},
Phys. Rev. A 67, 053613 
\hrefl{2003}{https://journals.aps.org/pra/abstract/10.1103/PhysRevA.67.053613}

\bibitem{Swallow4}
O. Fialko, M.-C. Delattre, J. Brand, A.R. Kolovsky, 
{\em Nucleation in Finite Topological Systems During Continuous Metastable Quantum Phase Transitions},
Phys. Rev. Lett. 108, 250402 
\hrefl{2012}{https://journals.aps.org/prl/abstract/10.1103/PhysRevLett.108.250402}

\bibitem{Swallow5}
S. Baharian, G. Baym, 
{\em Bose-Einstein condensates in toroidal traps: Instabilities, swallow-tail loops, and self-trapping},
Phys. Rev. A 87, 013619 
\hrefl{2013}{https://journals.aps.org/pra/abstract/10.1103/PhysRevA.87.013619}




%%% hysteresis and ring experiments 

%\cite{exprDimerHys}
%\cite{exprRingRev,exprRingNIST}

\bibitem{exprDimerHys} 
A.~Trenkwalder, G.~Spagnolli, G.~Semeghini, S.~Coop, M.~Landini, P.~Castilho, L.~Pezzè, G.~Modugno, M.~Inguscio, A.~Smerzi, M.~Fattori, 
{\em Quantum phase transitions with parity-symmetry breaking and hysteresis}, 
Nature Physics {12}, 826 
\hrefl{2016}{https://doi.org/10.1038/nphys3743}.

\bibitem{exprRingRev}
A.L.~Fetter, 
{\em Rotating trapped Bose-Einstein condensates}, 
Rev. Mod. Phys. {/bf 81}, 647 
\hrefl{2009}{https://doi.org/10.1103/RevModPhys.81.647}.

\bibitem{exprRingNIST}
S.~Eckel, J. G.~Lee, F.~Jendrzejewski, N.~Murray, C. W.~Clark,
C. J.~Lobb, W. D.~Phillips, M.~Edwards, G. K.~Campbell,
{\em Hysteresis in a quantized superfluid ‘atomtronic’ circuit}, Nature
(London) {506}, 200 
\hrefl{2014}{https://doi.org/10.1038/nature12958}.

%cst
\bibitem{cst}
K. Smith-Mannschott, M. Chuchem, M. Hiller, T. Kottos, D. Cohen, 
{\em Occupation Statistics of a BEC for a Driven Landau-Zener Crossing},
Phys. Rev. Lett. {102}, 230401 
\hrefl{2009}{http://dx.doi.org/10.1103/PhysRevLett.102.230401}  




%%% tunneling

\bibitem{dimerSplit}
G. Kalosakas, A.R. Bishop, and V.M. Kenkre, 
{\em Multiple-timescale quantum dynamics of many interacting bosons in a dimer},
J. Phys. B 36, 3233 
\hrefl{2003}{https://iopscience.iop.org/article/10.1088/0953-4075/36/15/305}


\end{thebibliography}

%%%%%%%%%%%%%%%%%%%%%%%%%%%%%%%%%%%%%%%%%%%%%%%%%%%%%%%%%%%%%%%%%%%%%%%%%%%%%%%%%%%%%%%%%%%%%%%%%%%%%%%%%%%%%%%%

%%%%%%%%%%%%%%%%%%%%%%%%%%%%%%%%%%%%%%%%%%%%%%%%%%%%%%%%%%%%%%%%%%%%%%%%%%%%%%%%%%%%%%%%%%%%%%%%%%%%%%%%%%%
%%%%%%%%%%%%%%%%%%%%%%%%%%%%%%%%%%%%%%%%%%%%%%%%%%%%%%%%%%%%%%%%%%%%%%%%%%%%%%%%%%%%%%%%%%%%%%%%%%%%%%%%%%%
\clearpage
\onecolumngrid
\pagestyle{empty}
\renewcommand\thelinenumber{}

\renewcommand{\thefigure}{S\arabic{figure}}
\setcounter{figure}{0}

\renewcommand{\theequation}{S-\arabic{equation}}
\renewcommand{\Eq}[1]{{\textcolor{blue}{Eq.}}~\!\!(\ref{#1})} 
\setcounter{equation}{0}

\Cn{
{\large \bf Quasistatic transfer protocols for atomtronic superfluid circuits} \\
{\large (Supplementary Material)}
} 

\Cn{
Yehoshua Winsten, Doron Cohen \\
{\footnotesize Department of Physics, Ben-Gurion University of the Negev, Beer-Sheva 84105, Israel}
}

%%%%%%%%%%%%%%%%%%%%%%%%%%%%%%%%%%%%%%%%%%%%%%%%%%%%%%%%%%%%%%%%%%%%%%%%%%%%%%%%%%%%%%%%%

%%%%%%%%%%%%%%%%%%%%%%%%%%%%%%%%%%%%%%%%%%%%%%%%%%%%%%%%%%%%%%%%%%%%%%%%%%%%%%%%%%%%%%%%%
\section{Energy landscape}

The SPs of the unperturbed Hamiltonian $\mathcal{H}^{(0)}(\varphi,n;M)$ for a given $M$ have to satisfy 
\beq
\frac{\partial \mathcal{H}^{(0)}}{\partial n} \ \ = \ \ \frac{\partial \mathcal{H}^{(0)}}{\partial \varphi} \ \ = \ \ 0
\eeq 
They are located, for any $M$, along ${\varphi=0,\pi}$, while $n$ should be determined from the equation  
\beq
-\frac{28}{9}n^4 
&+& \left( -4\frac{\mathcal{E}_{\parallel}}{NU} + \frac{14}{9} \right)n^3 
+ \left( \left(\frac{\mathcal{E}_{\parallel}}{NU}+\frac{1}{2}\right)^2+\frac{28}{9}\left(\frac{M}{N}\right)^2-\frac{4}{9} \right)n^2 
\\ \nonumber 
&+& \left(4\frac{\mathcal{E}_{\parallel}}{NU}+\frac{2}{9}\right)\left(\frac{M}{N}\right)^2 n 
-\left(\left(\frac{\mathcal{E}_{\parallel}}{NU}+\frac{1}{2}\right)^2+\frac{16}{9}\left(\frac{M}{N}\right)^2\right)\left(\frac{M}{N}\right)^2 \ \ =  \ \ 0
\eeq 
In the equation above $n$ is the normalized occupation, namely ${n:=n/N}$.
This equation has 4 roots, and at most two of them are within the physical range ${n\in[0,1]}$.
The central SP is ${n=0}$ for ${M=0}$.

\begin{figure}
	\centering
	(a) \hspace*{16cm}  \\ \vspace*{-4mm}
	\includegraphics[width=4.3cm]{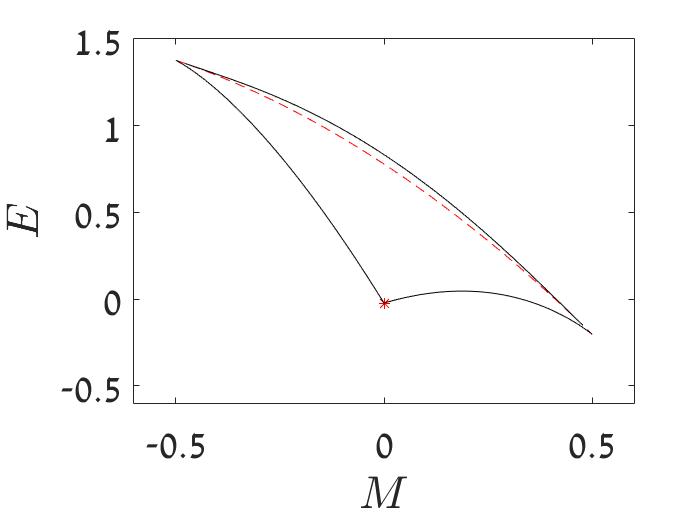}
	\includegraphics[width=4.3cm]{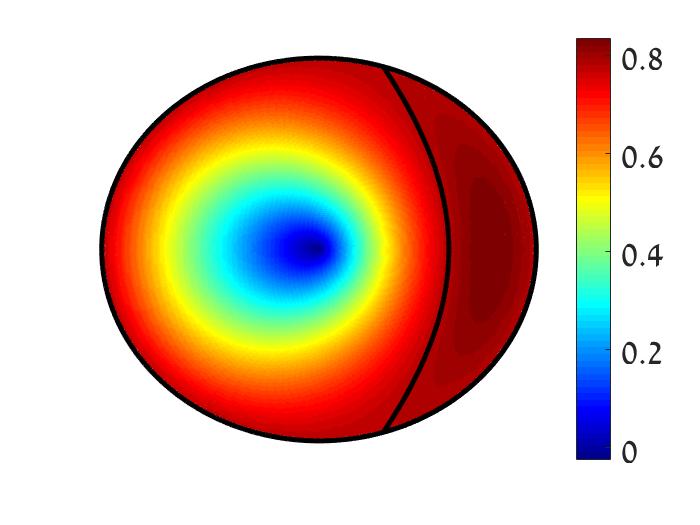} 
	\includegraphics[width=4.3cm]{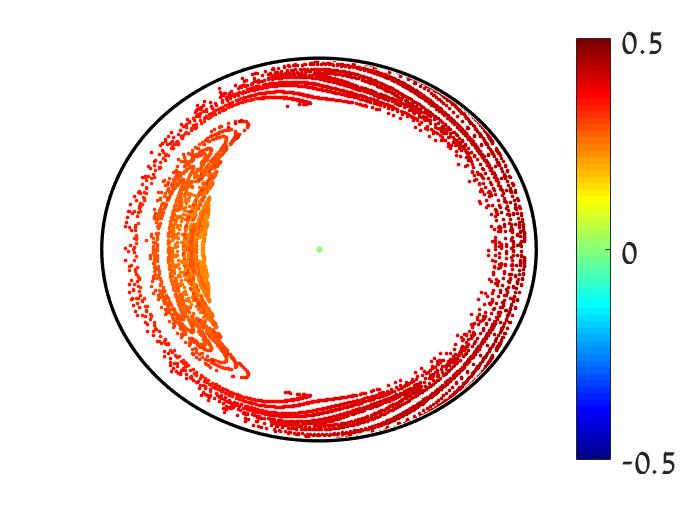} 
	\\
	(b) \hspace*{16cm}  \\ \vspace*{-4mm}
	\includegraphics[width=4.3cm]{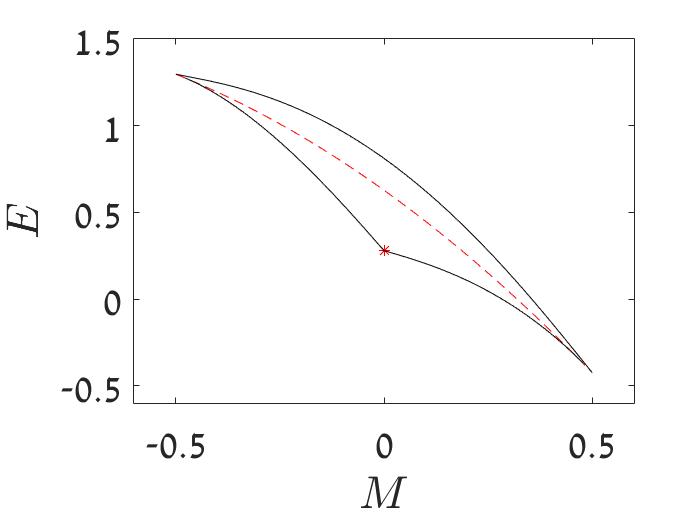}
	\includegraphics[width=4.3cm]{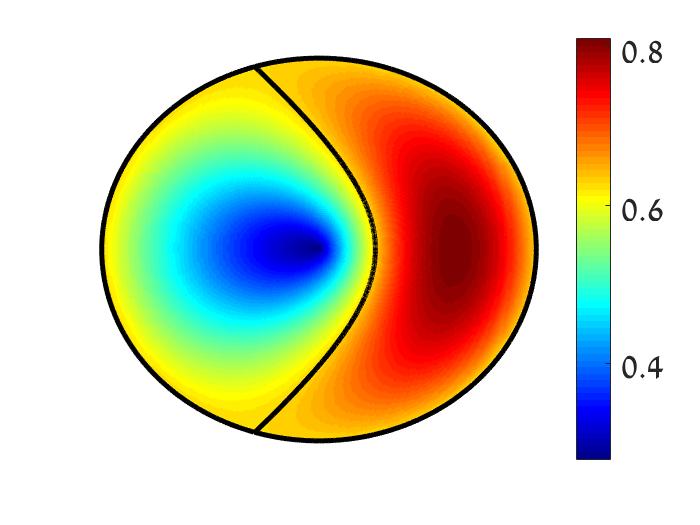} 
	\includegraphics[width=4.3cm]{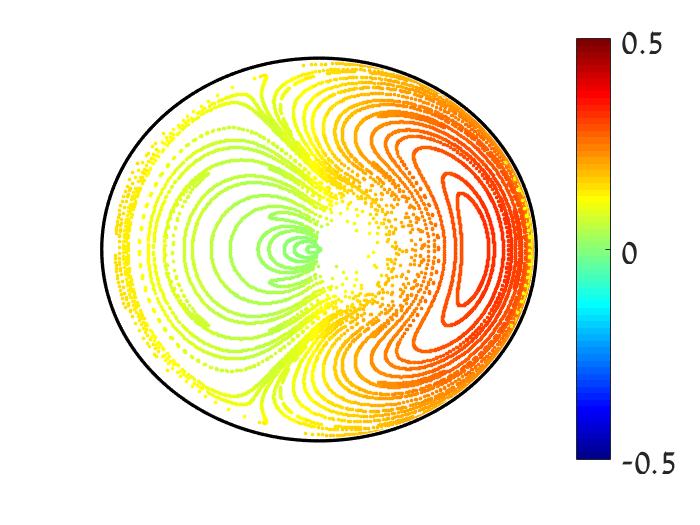} 
	\\
	(c) \hspace*{16cm}  \\ \vspace*{-4mm}
	\includegraphics[width=4.3cm]{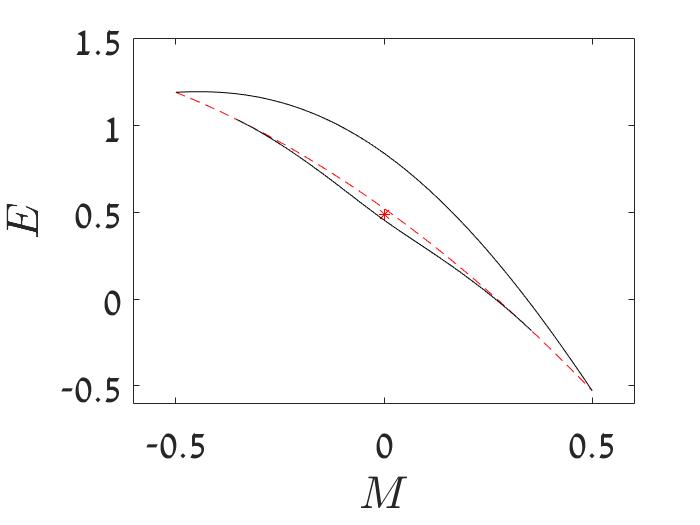}
	\includegraphics[width=4.3cm]{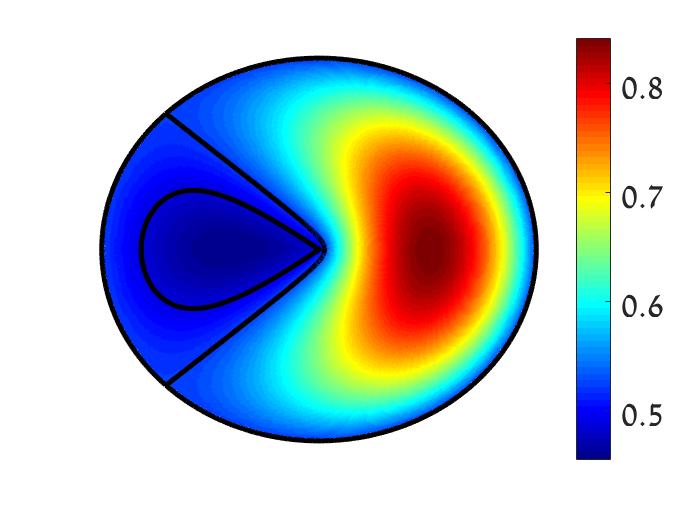} 
	\includegraphics[width=4.3cm]{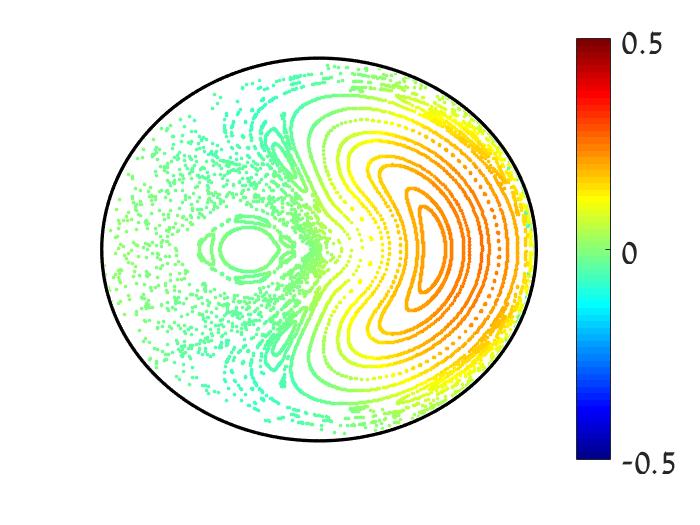} 
	\\
	(d) \hspace*{16cm}  \\ \vspace*{-4mm}
	\includegraphics[width=4.3cm]{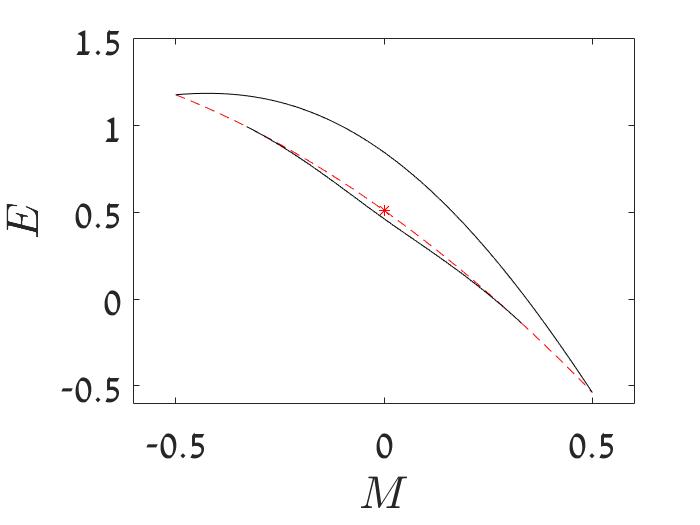}
	\includegraphics[width=4.3cm]{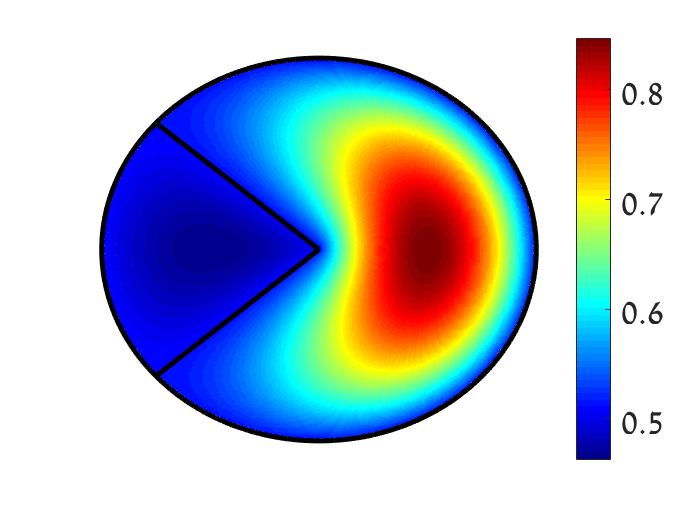} 
	\includegraphics[width=4.3cm]{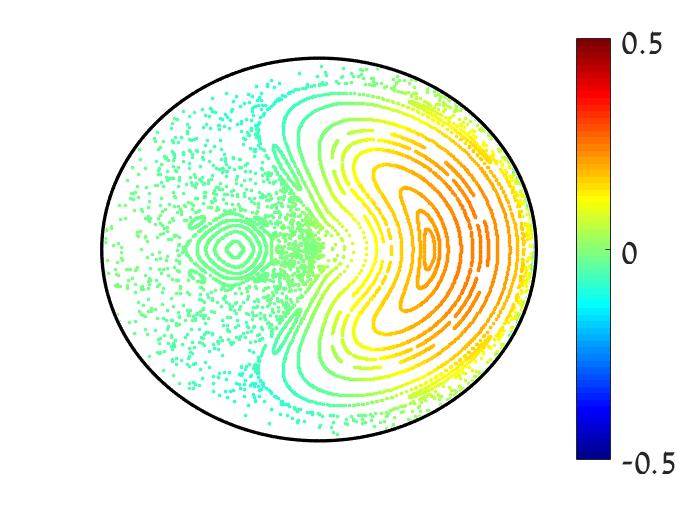} 
	\\
	(e) \hspace*{16cm}  \\ \vspace*{-4mm}
	\includegraphics[width=4.3cm]{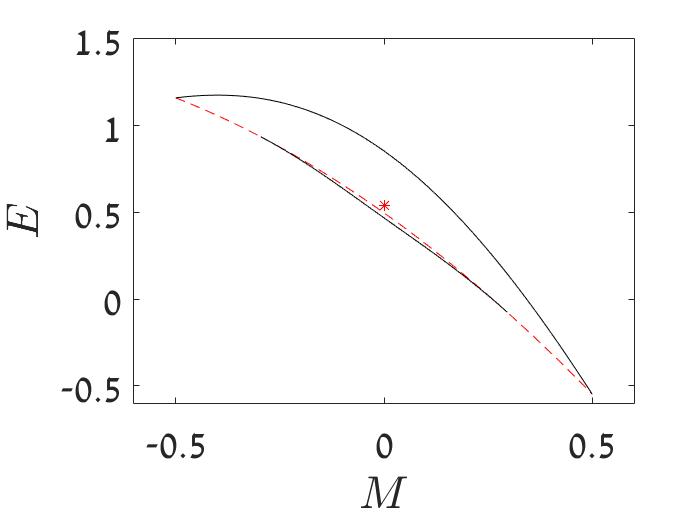}
	\includegraphics[width=4.3cm]{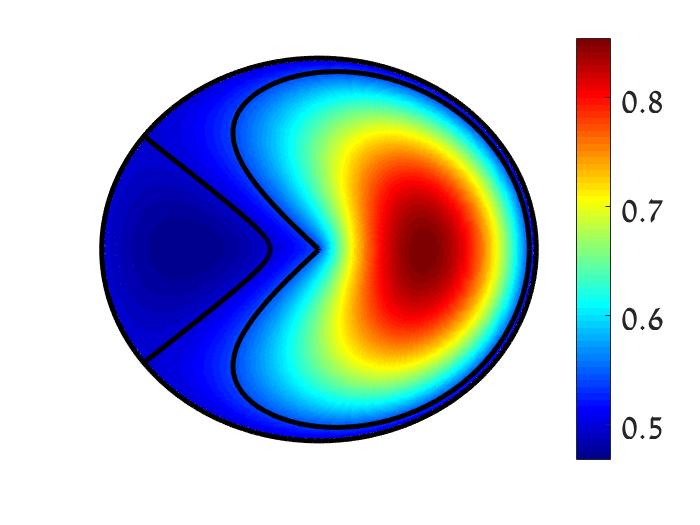} 
	\includegraphics[width=4.3cm]{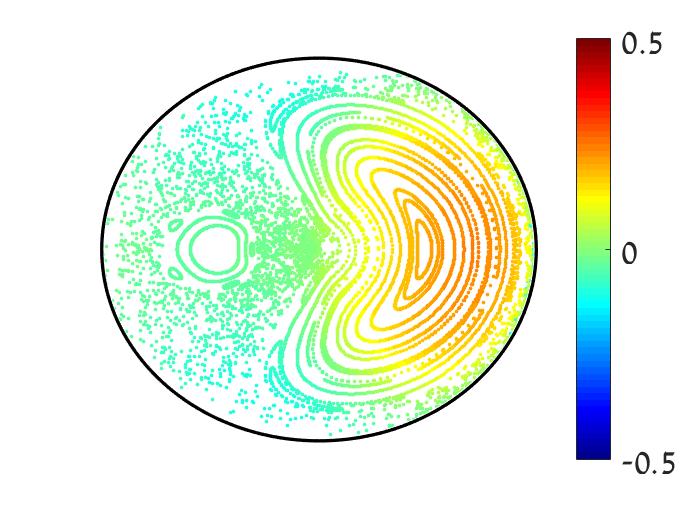} 
	\\
	(f) \hspace*{16cm}  \\ \vspace*{-4mm}
	\includegraphics[width=4.3cm]{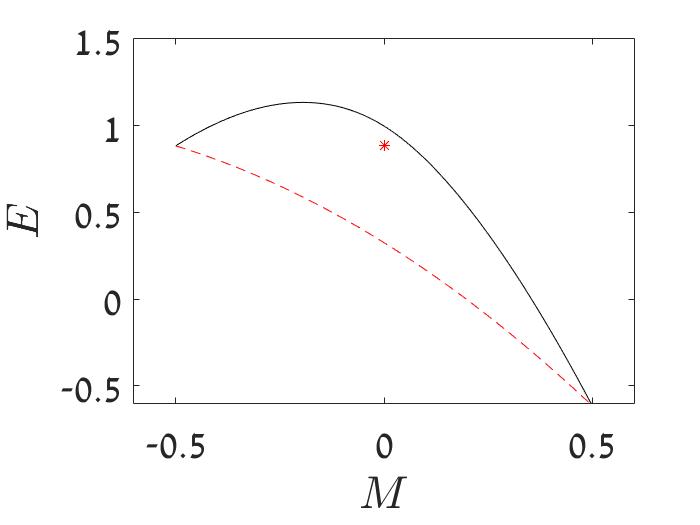}
	\includegraphics[width=4.3cm]{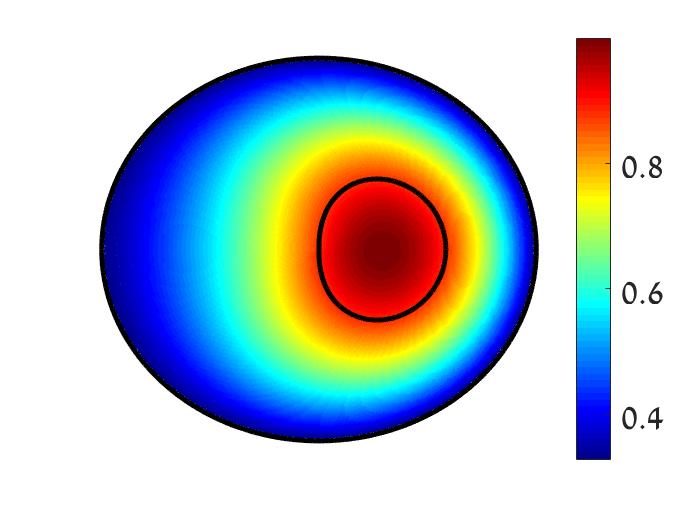} 
	\includegraphics[width=4.3cm]{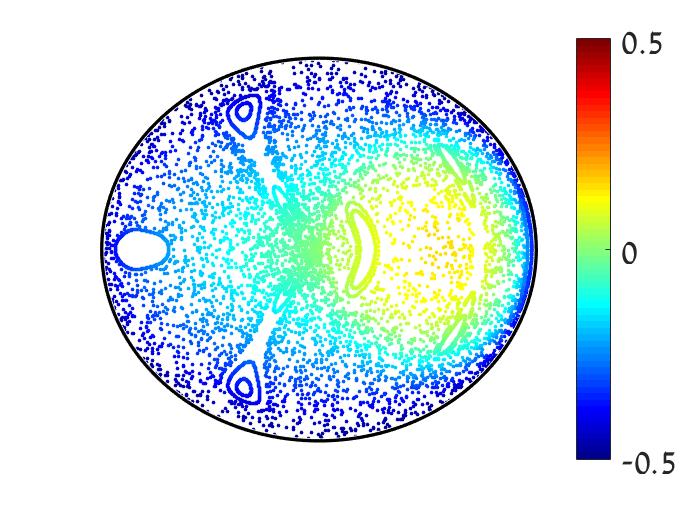} 
	\caption{\label{fLandscape}
		Left column: the energy landscape of $H^{(0)}$ for $u=2.3$. 
		Panels (a)-(f) are for $1.1\pi,1.4\pi,1.6\pi,\Phi_{\text{swp}},1.65\pi,2\pi$. 
		For each $M$ we find the floor (minimum) and the maximum of the energy, 
		and get the Black solid lines that bounds the spectrum from below and from above. 
		We also find for each $M$ the energy of the ${n=N/2}$ peripheral SP, and get the red dashed line.
		When the dashed line comes between the solid lines, it means that the peripheral SPs become saddles.
		The energy of the ${n=M=0}$ central SP is indicted by a red dot.
		When the red dot comes above the floor, it means that the central SP becomes an unstable saddle. 
		When the red dot crosses the dashed line, there is a swap of separatrices.
		At $\Phi_{\text{swp}}$ the two SPs are connected by a single level curve.      
		The middle column provides vertical section of the energy 
		landscape, namely ${E=\mathcal{H}^{(0)}(\varphi,n;M{=}0)}$.
		The right column displays Poincare sections at the central SP energy. 
		The trajectories are generated by $\mathcal{H}$ and their section-points are color-coded by $M$.
		Note that $M$ is not a constant of motion.} 
\end{figure}

The left column of \Fig{fLandscape} illustrates the energy landscape of $\mathcal{H}^{(0)}$ 
for representative values of $\Phi$. For each $M$ we find the floor (minimum) and the maximum of the energy, 
and get the Black solid lines that bound the spectrum from below and from above. 
In particular we indicated by a red point the energy $E_0$ of the central SP ($n{=}M{=}0$). 
Note that each point on the upper solid line is formally a peripheral SP (${n=N/2}$) 
of the unperturbed Hamiltonian for a given~$M$, which represents a totally depleted state.  
Explicit expressions for $E_0$ and for $E_{\infty}(M)$ are provide in \Eq{eE0} and \Eq{eEMinfty}.   
When the dashed line comes between the solid lines, it means that the peripheral SPs become saddles.
This happens in the range
\beq
3\arccos\left(\frac{1}{3}u\right) \ < \ \Phi \ < 3\arccos\left(-\frac{1}{9}u\right)
\eeq 
The central SP is the global minimum of the energy landscape up to $\Phi_{\text{mts}}$ of \Eq{PhiMTS}.
It is deduced from the equation ${ E_0 > E_{\infty}(N/2) }$.
For larger $\Phi$ the central SP it is still a local minimum, up to $\Phi_{\text{stb}}$ of \Eq{PhiSTB}. 
This value can be extracted from the Bogolyubov analysis: 
the SP becomes a saddle once $\omega_{-}$ of \Eq{eBog} changes sign and becomes negative.
When the red dot comes above the floor, see \Fig{fLandscape}c, 
it becomes dynamically unstable, 
and the Bogolyubov frequencies becomes complex.
This happens once we cross  $\Phi_{\text{dyn}}$ of \Eq{PhiDYN}. 
When the red dot crosses the dashed line (\Fig{fLandscape} panels c-d-e), 
it means that swap of separatrices takes place.
The transition happens when ${E_0=E_\infty(M{=}0)}$, leading to $\Phi_{\text{swp}}$ of \Eq{PhiSWP}. 
At the swap, the two SPs are connected by a single level curve. 
If the non-integrable terms $\mathcal{H}^{\pm}$ are included, 
this level curve becomes a chaotic strip. 
Thus a corridor is formed, that connects the central SP with the peripheral SPs.
This corridor remains open for a small range of $\Phi$ values around $\Phi_{\text{swp}}$.

For $\Phi{=}3\pi$ the ${n{=}0}$ central SP gets its highest value, 
which is not necessarily the maximum of the energy landscape. 
By the Bogolyubov analysis we can identify a critical value ${u_c=9/4}$. 
For large interaction (${u>u_c}$), as in \Fig{fLandscape}), 
the central SP is not the maximum of the landscape.
Rather, the new maxima support a self-trapped condensates.          
On the other hand, for weak interaction (${u<u_c}$), once we cross 
\beq
\Phi_{\text{dyn-end}} \ \ = \ \ 3\arccos{\left(-\frac{9}{4}u\right)}
\eeq
the central SP is stable again, and at $\Phi{=}3\pi$ it becomes a stable maximum.

The middle column of \Fig{fLandscape} provides vertical section of the energy 
landscape, namely ${E=\mathcal{H}^{(0)}(\varphi,n;M{=}0)}$.
The right column of \Fig{fLandscape} displays Poincare sections at the central SP energy. 
The trajectories are generated by $\mathcal{H}$ and their section-points are color-coded by $M$.
Note that $M$ is not a constant of motion. Quasi-regular trajectories tend to be mono-chromatic, 
while chaotic trajectories span a relatively wide range of $M$ values.

%%%%%%%%%%%%%%%%%%%%%%%%%%%%%%%%%%%%%%%%%%%%%%%%%%%%%%%%%%%%%%%%%%%%%%%%%%%%%%%%%%%%%%%%%
\clearpage

% Energy landscape (E,M) - panel for each Phi 

%%%%%%%%%%%%%%%%%%%%%%%%%%%%%%%%%%%%%%%%%%%%%%%%%%%%%%%%%%%%%%%%%%%%%%%%%%%%%%%%%%%%%%%%%
\clearpage
\section{Bogolyubov frequencies}

The Bogolyubov procedure brings the Hamiltonian in the vicinity of the SP to a diagonalized form.
\beq
\mathcal{H} \ \ \approx \ \ E[\text{SP}] + \sum_q \omega_q c_q^{\dag} c_q
\eeq

The equations of motion are: \ \  ${ \dot{z} = \mathbb{J} \partial \mathcal{H} }$ 

For one degree-of-freedom the canonical coordinates are \ ${ z = (a, \bar{a}) }$  
   
The symplectic matrix $\mathbb{J}$ is the second Pauli matrix. 

Hence an equivalent compact equation is:  \ \  ${ \dot{a} = -i \bm{h}[a,\bar{a}] a  }$

The SP satisfies ${\dot{a}=0}$, provided ${ \mathcal{H}:= \mathcal{H}-\mu N }$.

The Hessian (calculated at an SP): \ \  ${ \bm{H} \equiv \partial \partial \mathcal{H} }$

Linearized Hamiltonian: \ \ ${ \mathcal{H} \ \ \approx \ \ \frac{1}{2} \sum_{\mu.\nu} \bm{H}_{\mu,\nu} z_{\mu} z_{\nu} }$

Linearized equations: \ \   ${\dot{z} = [\mathbb{J} \bm{H}] \, z}$

Characteristic equation: \ \ ${\det(\lambda - \mathbb{J} \bm{H}) = 0}$

Eigenvalues are: ${\lambda_{q,\pm} = \pm i\omega_q}$ \ \   (one should be careful about the sign) 

One pair of frequencies is zero because the total occupation ($N$) is conserved.

\ \\

\sect{One site}
Consider one-site Hamiltonian ${ \mathcal{H} = \epsilon_0 \bar{a}a + \frac{U}{2} \bar{a} \bar{a} a a }$

Here ${ h[a,\bar{a}] =  \epsilon_0 + U \bar{a} a }$. 

The SP for $N$ particles is at ${a=\sqrt{N}}$ with ${\mu = \epsilon_0+NU }$. 

Accordingly the hessian at the SP is  
\beq
\bm{H} = \left(\amatrix{0 & \varepsilon_0 - \mu \cr \varepsilon_0 - \mu & 0 }\right)  
+ U\left(\amatrix{\bar{a}\bar{a} & 2\bar{a}a \cr 2\bar{a}a & aa }\right)_{a:=\sqrt{N}}
\eeq
The characteristic equation gives the trivial frequency ${\omega_0=0}$. 

\ \\

\sect{Ring}
For $M$ sites, the zero-momentum SP is associated with ${ \mu=\varepsilon_0+(NU/M) }$, and we get 
\beq
\bm{H} = \left(\amatrix{ 0 & \bm{h}_0-\mu  \cr \bm{h}_0-\mu & 0 }\right)
+ \frac{NU}{M} \left(\amatrix{ \bm{1} & \bm{2} \cr \bm{2} & \bm{1} }\right)
\eeq
where ${\bm{h}_0}$ is the kinetic part of $\bm{h}[a,\bar{a}]$ (only hopping terms, no interaction), 
and $\bm{1}$ (identity) and $\bm{2}$ (twice the identity) are $M \times M$ diagonal matrices (reflect the interactions). 
Note that the $\bm{2}$ can be absorbed into the kinetic matrix~${\bm{h}_0}$, 
while the $\bm{1}$ elements are related to terms of the type $a_j a_j$.   
Switching to the momentum basis the kinetic matrix becomes diagonal, while  
\beq
\bm{1} \mapsto \left(\amatrix{1 & 0 & 0 \cr 0 & 0 & 1 \cr 0 & 1 & 0 }\right)
\eeq
The above matrix includes the $k=0$ block plus one representative $(k,-k)$ block.  
If we look on $\mathbb{J} \bm{H}$, we see that it decouples into blocks.
All the block has the structure $\Omega_z\bm{\sigma}_z + i\Omega_y\bm{\sigma}_y$, up to a constant.
Note that $\Omega_z = \Omega_y$ is an exceptional point with zero eigenvalues.  
Indeed the $k=0$ block provides the zero frequencies, 
and the other blocks (without the $-i$ prefactor) are
\beq
\left(\amatrix{ \mathcal{E}_{k} & 0  \cr 0 &  -\mathcal{E}_{-k}  }\right)
+ \frac{NU}{M} \left(\amatrix{ 0 & 1 \cr -1 & 0 }\right)
\eeq
where ${\mathcal{E}_{k} = \varepsilon_k -\varepsilon_0 + (NU/M)}$. 
Note that the block with ${k\mapsto -k}$ provides frequencies with opposite signs. 
We conclude that 
\beq
\omega_{q,\pm} \ = \ \pm \left(\frac{\mathcal{E}_{q} - \mathcal{E}_{-q}}{2} \right) 
+ \sqrt{ \left(\frac{\mathcal{E}_{q} + \mathcal{E}_{-q}}{2}\right)^2 - \left(\frac{NU}{M}\right)^2 }
\eeq
The correctness of the sign convention can be tested by setting $U=0$.  

\begin{figure}[b!]
\centering
\includegraphics[width=7cm]{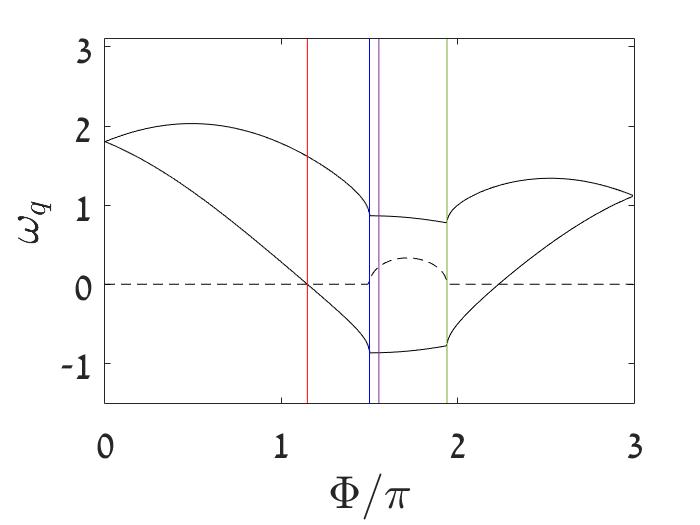}  
\includegraphics[width=7cm]{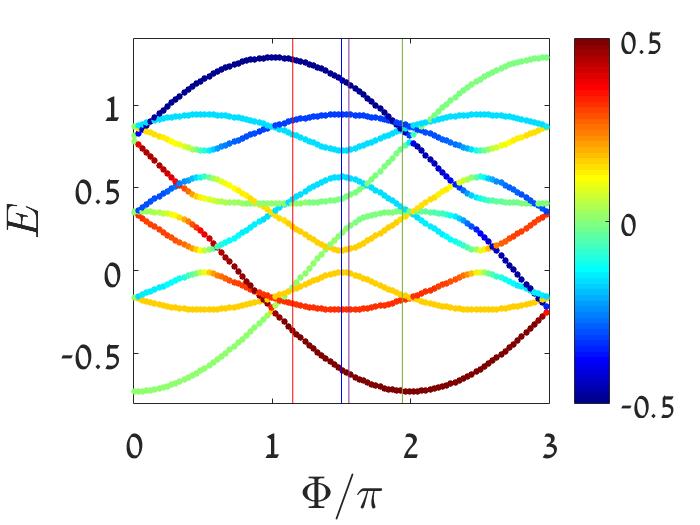}
\\
\includegraphics[width=7cm]{BG2-3}  
\includegraphics[width=7cm]{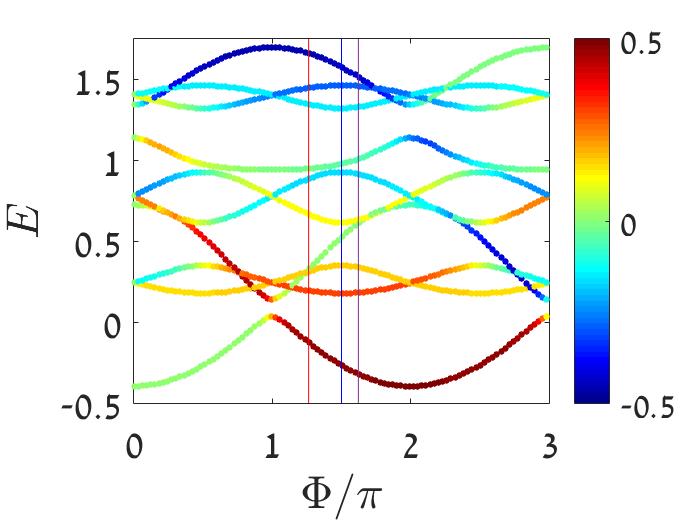}
\\
\includegraphics[width=7cm]{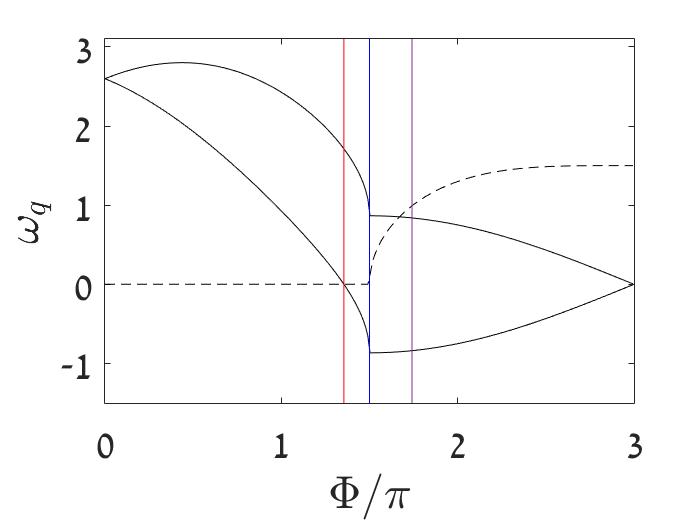}  
\includegraphics[width=7cm]{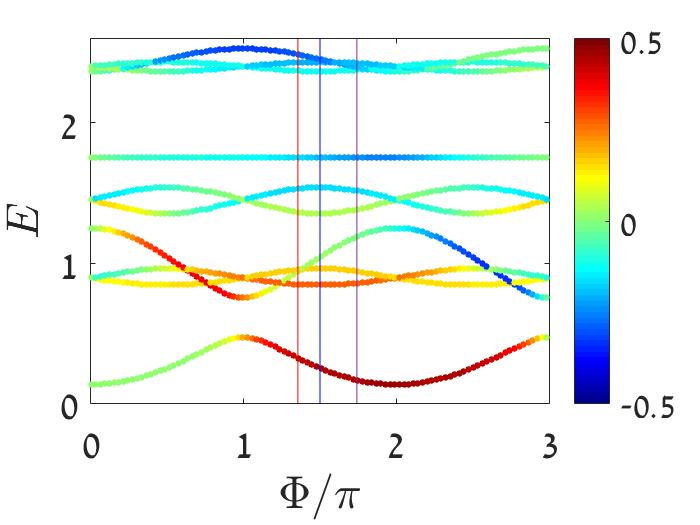}
\caption{{\bf Parametric variation of the energy landscape.} 
Left: The Bogolyubov frequencies for a ${k=0}$ condensate. 
The vertical lines from left to right are for $\Phi_{\text{stb}}$, $\Phi_{\text{dyn}}$ and $\Phi_{\text{swp}}$.
Right: The many body energy levels $E_n$ for $N=3$ particles as a function of $\Phi$. 
The points are color-coded by the expectation value of $M$.  
The calculations are done from up to down for ${u=1.0,2.3,4.5}$. 
In the first row (weak interaction) also $\Phi_{\text{dyn-end}}$ is indicated. 
}\label{fig:BG}
\end{figure}

The Bogolyubov frequencies are calculated as a function of $\Phi$ in \Fig{fig:BG}.
The implications of the various crossovers are reflected in the parametric 
diabatic evolution of the $E_0$ level in the quantum spectrum (right panels).

%%%%%%%%%%%%%%%%%%%%%%%%%%%%%%%%%%%%%%%%%%%%%%%%%%%%%%%%%%%%%%%%%%%%%%%%%%%%%%%%%%%%%%%%%%%%%%%%%%%%%%%%%%%%%%%
%\clearpage

% Energy levels for u=1.0, 2.3, 4.5
% Bogo for the same plot

%%%%%%%%%%%%%%%%%%%%%%%%%%%%%%%%%%%%%%%%%%%%%%%%%%%%%%%%%%%%%%%%%%%%%%%%%%%%%%%%%%%%%%%%%%%%%%%%
%%%%%%%%%%%%%%%%%%%%%%%%%%%%%%%%%%%%%%%%%%%%%%%%%%%%%%%%%%%%%%%%%%%%%%%%%%%%%%%%%%%%%%%%%%%%%%%%
%%%%%%%%%%%%%%%%%%%%%%%%%%%%%%%%%%%%%%%%%%%%%%%%%%%%%%%%%%%%%%%%%%%%%%%%%%%%%%%%%%%%%%%%%%%%%%%%
%%%%%%%%%%%%%%%%%%%%%%%%%%%%%%%%%%%%%%%%%%%%%%%%%%%%%%%%%%%%%%%%%%%%%%%%%%%%%%%%%%%%%%%%%%%%%%%%
\clearpage
\end{document}